\documentclass[fleqn,usenatbib]{mn2e}

\usepackage{aas_macros}
\usepackage{color}
\usepackage{bm}
\usepackage[colorlinks=true, allcolors=blue]{hyperref}
\usepackage{newtxtext,newtxmath}

\usepackage[T1]{fontenc}
\usepackage{ae, aecompl}
\usepackage{times}
\usepackage{amssymb}
\usepackage{float}
\usepackage{graphicx}
\usepackage{epstopdf}
\usepackage{lscape}
\usepackage{threeparttable}
\usepackage{bigints}
\usepackage{soul}

\newcommand{\kms}{km s$^{-1}$}
\newcommand{\Gaia}{{\it Gaia}}

\defcitealias{Marchetti+19}{M19}

\title[Unbound stars in {\Gaia} EDR3]{{\Gaia} EDR3 in 6D: Searching for unbound stars in the Galaxy}

\author[Tommaso Marchetti]{Tommaso Marchetti\thanks{E-mail: \href{mailto:tommaso.marchetti@eso.org}{tommaso.marchetti@eso.org}}\\
European Southern Observatory, Karl-Schwarzschild-Strasse 2, 85748 Garching bei M{\"u}nchen, Germany \\
}

\date{Accepted XXX. Received YYY; in original form ZZZ}

\begin{document}

\pagerange{\pageref{firstpage}--\pageref{lastpage}} \pubyear{2020}

\maketitle

\label{firstpage}

\begin{abstract}

The early third data release (EDR3) of the European Space Agency satellite {\Gaia} provides coordinates, parallaxes, and proper motions for $\sim 1.47$ billion sources in our Milky Way, based on 34 months of observations. The combination of {\Gaia} DR2 radial velocities with the more precise and accurate astrometry provided by {\Gaia} EDR3 makes the best dataset available to search for the fastest nearby stars in our Galaxy. We compute the velocity distribution of $\sim 7$ million stars with precise parallaxes, to investigate the high-velocity tail of the velocity distribution of stars in the Milky Way. We release a catalogue with distances, total velocities, and corresponding uncertainties for all the stars considered in our analysis\footnotemark.
By applying quality cuts on the {\Gaia} astrometry and radial velocities, we identify a clean subset of 94 stars with a probability $P_\mathrm{ub}> 50 \%$ to be unbound from our Galaxy. 17 of these have $P_\mathrm{ub}> 80\%$ and are our best candidates. We propagate these stars in the Galactic potential to characterize their orbits. We find that 11 stars are consistent with being ejected from the Galactic disk, and are possible hyper-runaway star candidates. The other 6 stars are not consistent with coming from a known star-forming region. We investigate the effect of adopting a parallax zero point correction, which strongly impacts our results: when applying this correction, we identify \emph{only} $12$ stars with $P_\mathrm{ub}> 50\%$, $3$ of these having $P_\mathrm{ub}> 80\%$. Spectroscopic follow-ups with ground-based telescopes are needed to confirm the candidates identified in this work.

\end{abstract}

\footnotetext{\url{https://sites.google.com/view/tmarchetti/research}}


\section{Introduction}
\label{sec:intro}

The majority of nearby stars rotate coherently with the Sun around the Galactic Centre in the shape of a flattened disk \citep{Oort27} with a velocity $\sim 240$ {\kms} \citep{bland-hawthorn+16}. The halo of the Milky Way, a diffuse stellar component extending far beyond the disk and showing large random motions, exhibits a typical velocity dispersion $\sim 150$ {\kms} \citep[e.g.][]{smith+09, evans+16}. Observations reveal the presence of nearby objects with motions deviating significantly from these components: high velocity stars, moving remarkably faster than the surrounding neighbours. These objects are unique tools to probe our Milky Way: they travel through the Galaxy during their lifetime, being observed far away from their birthplace, and they give us insights into some of the most extreme phenomena in the local Universe. Fast stars have been observed across the whole velocity spectrum, up and beyond the escape speed from the Galaxy, which decreases from $\sim 530$ {\kms} at the Sun position \citep{deason+19} to $\sim 380$ \kms in the outer halo \citep{williams+17}. 

Historically, two main ejection mechanisms have been introduced to explain the puzzling observations of young O- and B-type stars at high Galactic latitudes, the so-called \emph{runaway stars} \citep{blaauw61}. The first scenario involves the explosion of a supernova in a binary system, with the companion star being ejected with a high velocity \citep{blaauw61, portegieszwart00}. The other possible explanation involves close encounters in dense stellar systems \citep{poveda+67}. Both scenarios have been shown to be responsible for the observed population of runaway stars, even if the relative contribution given by these mechanisms is still unclear \citep{hoogerwerf+01, silva+11}. The majority of runaway stars produced by both scenarios have typical velocities below $\sim 10$ {\kms} \citep{portegieszwart00,eldridge+11, Renzo+19}, the so-called \emph{walkaway stars} \citep{deMink+14}. Highest velocities up to several hundreds of {\kms} can be attained for particular combinations of the parameter space \citep[e.g.][]{leonard+90, portegieszwart00, przybilla+08, gvaramadze+09, Renzo+19}, and predicted upper limits are around $450$ {\kms} for early type stars \citep{Irrgang+19}. Runaway stars of lower masses can be ejected with unbound velocities \citep[][]{Leonard91, tauris15}, but the rate of ejection of unbound runaway stars (also called \emph{hyper-runaway stars}) is predicted to be extremely small \citep{perets12, brown15, Evans+20}.

Extremely high velocities of the order of thousands of {\kms} can instead be easily reached by \emph{hypervelocity stars} (HVSs), first predicted by \cite{hills88} as the result of a close interaction between a binary system and the massive black hole residing in the Galactic Centre \citep[see][for a comprehensive review on HVSs]{brown15}. The first HVS candidate was observed by \cite{brown+05} while targeting blue horizontal branch stars in the outer halo of the Milky Way. This star has a total velocity of $673$ {\kms}, which is high enough to escape from the gravitational field of the Galaxy \citep{Brown+14}. A following dedicate survey, the MMT HVS Survey, discovered a total of 21 late B-type HVSs, unbound from radial velocity alone \citep{Brown+14, brown+15}. The fastest HVS has been recently discovered by \citet{Koposov+20}: S5-HVS1, an A-type star at a distance of $\sim 9$ kpc from the Sun, with a total velocity of $\sim 1700$ {\kms}. The velocity vector of S5-HVS1 points unambiguously away from the Galactic Centre, making the Hills' mechanism the only possible explanation for its extreme acceleration. The unique property of HVSs is that, given their unbound velocities, they can travel hundreds of kpc during their lifetime, with their orbit being shaped by the underlying Galactic potential. HVSs can therefore be used as tools to probe the shape and the orientation of the matter (and dark matter) distribution in our Galaxy \citep[e.g.][]{gnedin+05, yu&madau07, Contigiani+19}, even if tight constraints are hampered by the precision of currently available data \citep{Rossi+17}. The same mechanism producing HVSs is expected to eject a population of \emph{bound} HVSs, with velocities sufficiently high to escape the gravitational field of the massive black hole, but not to exceed the escape velocity from the Galaxy \citep[e.g.][]{bromley+06, kenyon+08}. The ejection velocity distribution in the Galactic Centre is expected to peak at bound velocities \citep[e.g.][]{rossi+14}, and bound candidates are expected to outnumber unbound HVSs from the inner bulge to $\sim 10$ kpc beyond the Solar neighbourhood \citep{kenyon+14, marchetti+18}.

Other ejection mechanisms have been introduced to explain the observations of stars exceeding the known limits for runaway stars. These include, but are not limited to, the ejection of runaway stars and HVSs from the Large Magellanic Cloud \citep[LMC,][]{boubert+16, boubert+17a}, the acceleration of single stars following the interaction between a black hole binary made of Sagittarius A$^*$ and an intermediate mass black hole \citep{yu&tremaine03, sesana+06, Rasskazov+19}, tidal interactions of dwarf galaxies with the Milky Way \citep{abadi+09}, and the infall of globular clusters towards the Galactic Centre \citep{capuzzodolcetta+15, fragione+16}. Relative contributions from the different mechanisms are still unknown.

The European Space Agency (ESA) satellite {\Gaia} has been scanning the sky in the Lagrangian point L2 since 2014, with the goal of producing the largest and most precise three-dimensional map of the Milky Way \citep{gaia}.
The most recent {\Gaia} data release is the early third data release (EDR3), out on 3 December 2020 \citep{gaiaedr3}. {\Gaia} EDR3 contains positions and magnitudes in the $G$ band for $\sim 1.81$ billion sources, and parallaxes and proper motions for a subset of $\sim 1.47$ billion sources \citep{Lindegren+20a}. {\Gaia} EDR3 provides also photometry in the Blue Pass (BP) and Red Pass (RP) filters for $\sim 1.55$ million objects \citep{Riello+20}. While no new radial velocities are provided as part of {\Gaia} EDR3 \citep{Seabroke+20}, and the catalogue content is similar to the one provided by the second {\Gaia} data release in April 2018 \citep[DR2,][]{gaiadr2}, the extended observational baseline of $34$ months (compared to the $22$ months of {\Gaia} DR2) results into more precise astrometric measurements. In particular, parallax errors have reduced by a factor $\sim 20\%$, and proper motions are twice more precise. In addition to the increased astrometric precision, also an overall reduction of systematics has been achieved. The parallax offset inferred from distant quasars has decreased from $-29$ $\mu$as in {\Gaia} DR2 to $-17$ $\mu$as in {\Gaia} EDR3 \citep{Lindegren+18, Lindegren+20b}. \citet{Lindegren+20b} provides a tentative formula to correct {\Gaia} EDR3 parallaxes, as a function of the position, photometry and astrometry of the source. {\Gaia} DR2 provides radial velocities for $\sim 7.2$ million bright stars ($G \lesssim 13$) with effective temperatures in the range $[3550, 6900]$ K \citep{Katz+19}, obtained with the Radial Velocity Spectrometer (RVS) on board of the satellite \citep{Cropper+18}. \citet{Boubert+19} introduced several quality cuts to select reliable radial velocities from {\Gaia} DR2, especially in crowded fields, where nearby bright stars might contaminate the radial velocity measurements. In {\Gaia} EDR3, $\sim 4000$ erroneous radial velocities from {\Gaia} DR2 have been excluded, including all the cases in which the absolute value of the radial velocity was above $625$ {\kms} \citep{Seabroke+20}. In addition, $\sim 10000$ radial velocities from {\Gaia} DR2 could not be associated to {\Gaia} EDR3 sources \citep{Seabroke+20}. New information, including new radial velocities, astrophysical parameters, spectra and non-single stars, will be included as part of the full third {\Gaia} data release (DR3), currently expected in the first half of 2022\footnote{\url{https://www.cosmos.esa.int/web/gaia/release}, accessed on January 28 2021}.

Several papers used the precise astrometry (in particular proper motions) from {\Gaia} DR2 to revisit our knowledge on the fastest stars in our Galaxy. \citet{Boubert+18} computed total velocities for all the previous known unbound star candidates, finding that the majority of the high velocity late-type stars are likely to be bound to the Milky Way. \cite{Brown+18} used the new {\Gaia} DR2 proper motion to test the origin of the known HVS candidates, confirming that the fastest stars are still consistent with coming from the Galactic Centre. \citet{erkal+19} traced back the orbit of one HVS candidate, HVS 3, to the centre of the LMC, providing evidence for the existence of a massive black hole with a mass of at least $4 \cdot 10^3 - 10^4$ M$_\odot$. \citet{marchetti+18} predicted the expected population of HVSs in {\Gaia}, finding that, even if thousands of HVSs with precise proper motions are already included in the {\Gaia} catalogue, the missing radial velocity measurement for the majority of them makes their identification extremely challenging.

The high velocity tail of the velocity distribution of stars with full phase space information in {\Gaia} DR2 has been characterized by several works. \citet[][hereafter \citetalias{Marchetti+19}]{Marchetti+19} derived distances and total velocities for all the $\sim 7.2$ million sources in {\Gaia} DR2 with full phase space information. The authors found $20$ stars with probabilities higher than $80 \%$ of being unbound from the Galaxy. Surprisingly, the majority of these are on orbits which are not pointing away from a known Galactic star forming region (such as the Galactic disk), but an extragalactic origin is preferred. \citet{Bromley+18} focused on a subset of the same sample, finding $2$ stars with a probability $\sim 100\%$ of being unbound from the Galaxy. In addition, the authors found $19$ stars with precise parallaxes and proper motions that are unbound in tangential velocity alone. \citet{hattori+18b} reported the discovery of $30$ old, metal-poor stars with total velocities higher than $480$ {\kms}, and provided constraints on the mass of the Milky Way, under the assumption that these stars are bound to the Galaxy. In this work, we update the results found in \citetalias{Marchetti+19}, revisiting the high velocity tail of the velocity distribution of stars with {\Gaia} DR2 radial velocities, using the new, precise astrometry from {\Gaia} EDR3. We derive distances and total velocities for all the stars with precise and positive parallaxes, and we trace-back in the Galactic potential the stars with highest probabilities of being unbound.

This paper is organized as follows. In Section \ref{sec:method} we detail our method to derive distances and velocities for the stars with full phase space information and precise parallaxes from {\Gaia} DR2 and {\Gaia} EDR3. In Section \ref{sec:vel_distr} we show the spatial and velocity distributions for the whole sample of stars, and we identify a clean subset of high velocity stars with accurate {\Gaia} measurements and high probabilities of being unbound from the Galaxy. Next, in Section \ref{sec:orbits} we trace-back in the Galactic potential the candidates with the highest probability of being unbound from the Galaxy, to characterize their orbits and to identify their ejection location. We introduce individual high velocity star candidates in Section \ref{sec:high_vel}. In Section \ref{sec:zp} we correct {\Gaia} EDR3 parallaxes using the zero point correction introduced in \citet{Lindegren+20b}, and we explore its impact on the high velocity star candidates identified in this work. Finally, we discuss our findings and present our conclusions in Section \ref{sec:discuss}. 

\section{Deriving distances and total velocities}
\label{sec:method}

\begin{figure}
	\centering
 	\includegraphics[width=\columnwidth]{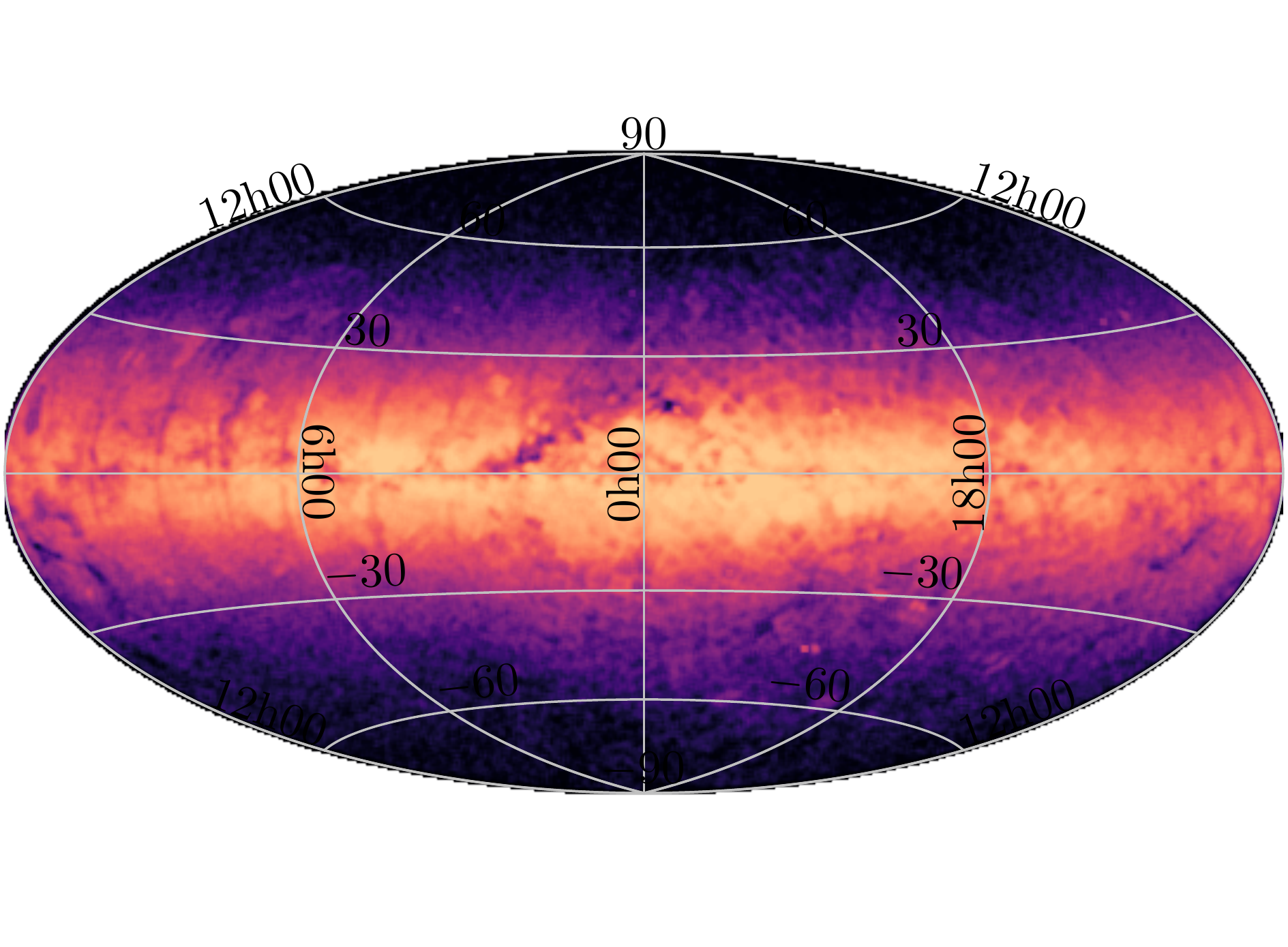}
	\caption{Aitoff projection of the sky density in Galactic coordinates of the 7209831 sources in {\Gaia} EDR3 with radial velocities from {\Gaia} DR2. We histogram-equalised the colourmap to increase the density contrast.}
	\label{fig:sky}
\end{figure}

{\Gaia} EDR3 provides right ascension $\alpha$, declination $\delta$, parallax $\varpi$, and proper motions $\mu_{\alpha*}$ and $\mu_\delta$ (the full 5-parameters astrometric solution\footnote{In {\Gaia} EDR3, a distinction has been made between the 5-parameters and the 6-parameters astrometric solutions \citep{Lindegren+20a}. The extra parameter is the astrometric pseudo-colour, which is determined for 882328109 {\Gaia} EDR3 sources. 238890 of the 7209831 {\Gaia} EDR3 sources  with radial velocities from {\Gaia} DR2 have a 6-parameters solution ($\sim 3\%$). In this paper we do not distinguish between the two types of astrometric solutions, and we make use of the pseudo-colour only in Section \ref{sec:zp}, when estimating the zero point in parallax for each source, following \citet{Lindegren+20b}.}) for 1467744818 stars \citep{gaiaedr3}. A  sample of 7209831 bright stars with {\Gaia} magnitudes $G \lesssim 13$ have radial velocities from {\Gaia} DR2 \citep{Seabroke+20}, and will be the main focus of this work. $6\%$ of these stars have a \textsc{source\_id} in {\Gaia} EDR3 different from the one reported in {\Gaia} DR2. The density of these sources, in Galactic coordinates, is shown in Fig. \ref{fig:sky}. We can clearly see the overdensity along the Galactic plane, with features related on large scales to dust structures and, on smaller scales, to the {\Gaia} scanning law and the RVS selection function. $7180466$ of these sources have also full astrometry from {\Gaia} EDR3. In this Section, we derive distances and velocities in the Galactocentric rest-frame for this sample of stars with full phase space information. We follow the same approach outlined in \citetalias{Marchetti+19}, which we report here again for clarity.

Deriving an accurate distance from an observed parallax is straightforward as $d=1/\varpi$ only when $\varpi > 0$ and $f \equiv \sigma_\varpi/\varpi \lesssim 20 \%$ \citep{bailer-jones, luri+18}, where $\sigma_\varpi$ is the reported random uncertainty in parallax. In this paper, we thus choose to focus only on the subset of 6969738 stars with precise parallaxes satisfying these criteria ($\sim 97\%$ of the total sample of stars with full phase space information). The sky density of this subset of stars does not differ appreciably from the one shown in Fig. \ref{fig:sky}. In principle, a Bayesian approach can be implemented to infer distances for stars with noisy or even negative parallaxes \citep[see][]{bailer-jones, astraatmadjaI, astraatmadjaII, bailer-jones+18, luri+18, bailer-jones+20}, but the resulting distance (and therefore total velocity) depends strongly on the choice of the prior probability on distances $P(d)$ \citep[e.g.][]{astraatmadjaII, marchetti+17}. Since our goal is to identify a clean, precise sample of high velocity stars in the Milky Way, we do not follow this approach, and we choose to restrict our analysis to the sources with precise parallaxes, for which we can get a clear identification of their trajectory. The remaining 210728 stars with either $\varpi < 0$ or $f > 20 \%$ are not considered further in this work. We note how the number of stars with positive and precise parallaxes increased from $\sim 89\%$ of the sample of stars with astrometry and radial velocities in {\Gaia} DR2, to $\sim 97\%$ in {\Gaia} EDR3.

For the subset of $\sim 7$ million stars with precise and positive parallaxes, we implement a Monte Carlo (MC) simulation to derive distances and velocities starting from the observables $(\alpha, \delta, \varpi, \mu_{\alpha*}, \mu_\delta, v_\mathrm{rad})$ and their corresponding uncertainties $\sigma_i$. We draw 5000 MC realizations of the astrometry of each star from a multi-variate Gaussian distribution, with mean vector:
\begin{equation}
\label{eq:mean}
    \mathbf{m} = [\mu_{\alpha*}, \mu_\delta, \varpi],
\end{equation}
and covariance matrix:
\begin{equation}
\label{eq:covmatr} 
\footnotesize
\arraycolsep=2pt
\thickmuskip =1.mu
\Sigma = {}
\left(
\begin{array}{@{}ccc@{}}
\sigma_{\mu_{\alpha *}}^2 & \sigma_{\mu_{\alpha *}} \sigma_{\mu_\delta} \rho(\mu_{\alpha *},\mu_\delta) & \sigma_{\mu_{\alpha *}} \sigma_\varpi \rho(\mu_{\alpha *},\varpi) \\
\sigma_{\mu_{\alpha *}} \sigma_{\mu_\delta} \rho(\mu_{\alpha *},\mu_\delta) & \sigma_{\mu_\delta}^2 & \sigma_{\mu_{\delta}} \sigma_\varpi \rho(\mu_\delta,\varpi) \\
\sigma_{\mu_{\alpha *}} \sigma_\varpi \rho(\mu_{\alpha *},\varpi) & \sigma_{\mu_\delta} \sigma_\varpi \rho(\mu_\delta,\mu_\varpi) & \sigma_\varpi^2 
\end{array}
\right),
\end{equation}
where $\rho(i,j)$ is the correlation coefficient between the two given astrometric parameters $i$ and $j$. In the equations above we have neglected uncertainties and correlations in the sky coordinates $\alpha$ and $\delta$, which are subdominant. Typical uncertainties in parallaxes are $\sim 0.02 - 0.03$ mas for the sample of bright stars considered in this paper, and typical errors in proper motions are $\sim 0.02 - 0.03$ mas yr$^{-1}$ \citep{gaiaedr3}. Radial velocities from the spectrometer on board of {\Gaia} are not correlated to the {\Gaia} EDR3 astrometry, so we draw the same number of MC radial velocity samples from a univariate Gaussian distribution centered on $v_\mathrm{rad}$ and with standard deviation equal to $\sigma_{v_\mathrm{rad}}$. Typical errors in {\Gaia} DR2 radial velocities are of the order of $\sim 200 - 300$ m s$^{-1}$ at the bright end, and of a few {\kms} at the faint end \citep{Katz+19}.

Distances are derived by inverting the MC parallax samples, $d = 1/\varpi$. To compute Cartesian coordinates and velocities in the Galactocentric frame, we assume a distance of the Sun from the Galactic Centre of $r_\odot = 8.122$ kpc \citep{Gravity+18}, a height of the Sun above the Galactic disk of $z_\odot = 25$ pc \citep{bland-hawthorn+16}, a circular velocity at the Sun's position of 235 {\kms} and a Sun's peculiar velocity vector $\mathbf{v_\odot} = [11.1, 12.24, 7.25]$ {\kms} \citep{schonrich09}.
We compute Galactic rectangular velocities $(U, V, W)$ following the convention that $U$ is positive when pointing toward the Galactic Centre, $V$ in the direction of Galactic rotation, and $W$ toward the North Galactic Pole \citep{johnson_soderblom}. The total velocity of each star in the Galactocentric frame is then obtained as $v_\mathrm{GC} = \sqrt{U^2 + V^2 + W^2}$. 

Finally, we estimate the probability $P_\mathrm{ub}$ of each star of being unbound from the Milky Way as the ratio of MC realizations resulting in a total velocity $v_\mathrm{GC}$ higher than the escape speed from the Galaxy at the position of the star. We compute the escape speed as a function of distance using the same Galactic potential model that we use for the orbits integration, as detailed in Section \ref{sec:orbits}. The resulting value at the Sun's position is $\sim 560$ {\kms}, consistent within the uncertainties with the estimates from \cite{smith+07}, \citet{williams+17} and \cite{monari+18}, and slightly higher than the value reported in \cite{deason+19}.

In Section \ref{sec:zp}, we discuss the impact of adopting a zero point correction to {\Gaia} EDR3 parallaxes, following the approach outlined in \citet{Lindegren+20b}. A comparison between the distances derived in this work and those recently inferred in \citet{bailer-jones+20} for $\sim 1.47$ billion stars in {\Gaia} EDR3 is presented in Appendix \ref{appendix:distances}.

\section{Spatial and velocity distributions}
\label{sec:vel_distr}

\begin{figure}
	\centering
 	\includegraphics[width=\columnwidth]{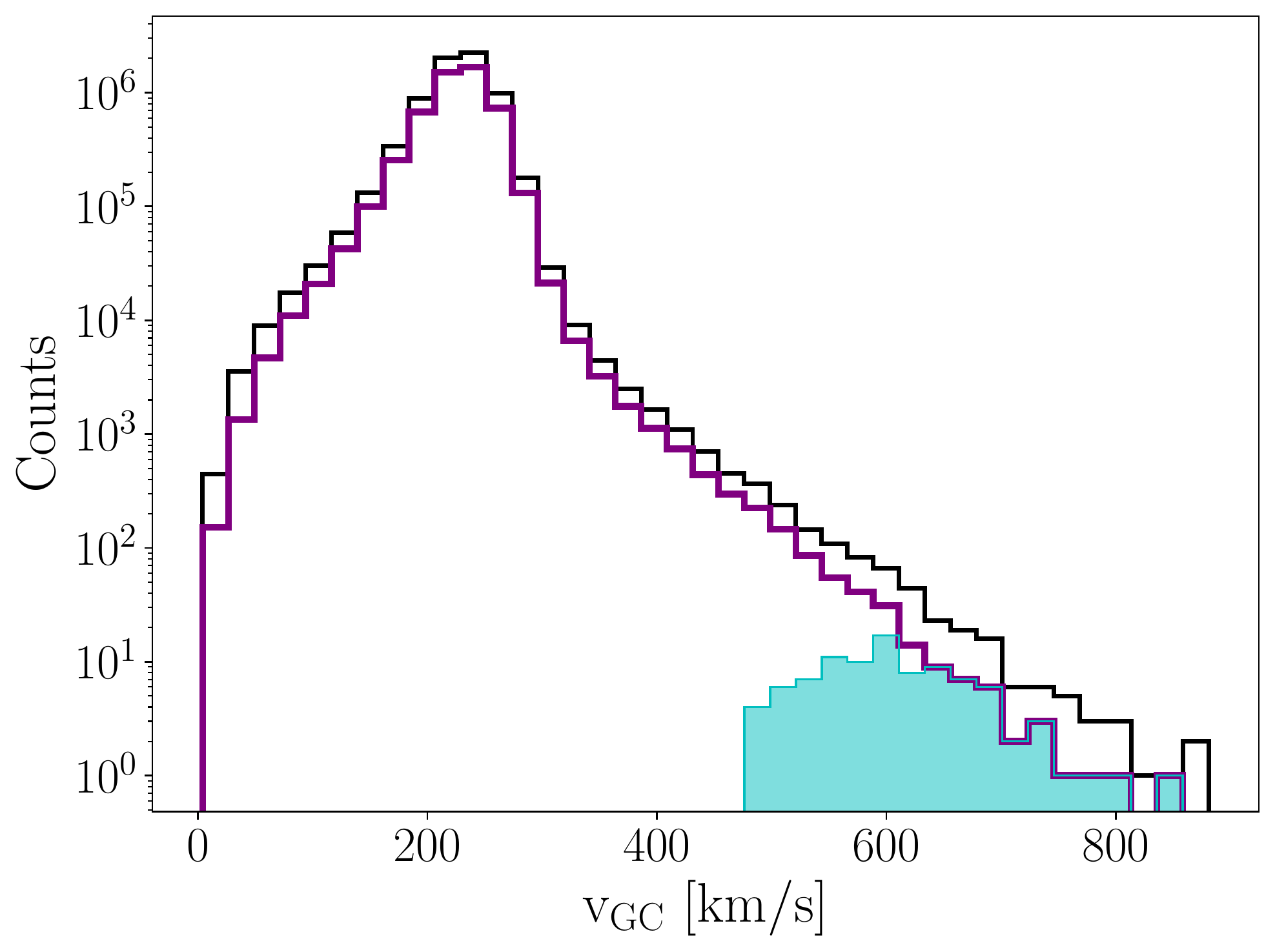}
	\caption{Histogram of median total velocities in the Galactic frame for all the 6969738 stars with precise and positive parallaxes considered in this paper (black line). The purple line shows the 5191458 objects surviving the quality cuts \textsc{ruwe} $<1.4$, \textsc{dr2\_rv\_nb\_transits} $ \geq 4$ and $\sigma_{v_\mathrm{GC}} / v_\mathrm{GC} < 30 \%$. The light blue histogram shows the distribution for the clean subset of 94 stars with $P_\mathrm{ub} > 0.5$, as discussed in Section \ref{sec:cuts}.}
	\label{fig:vGC_hist}
\end{figure}

\begin{figure*}
	\centering
 	\includegraphics[width=\textwidth]{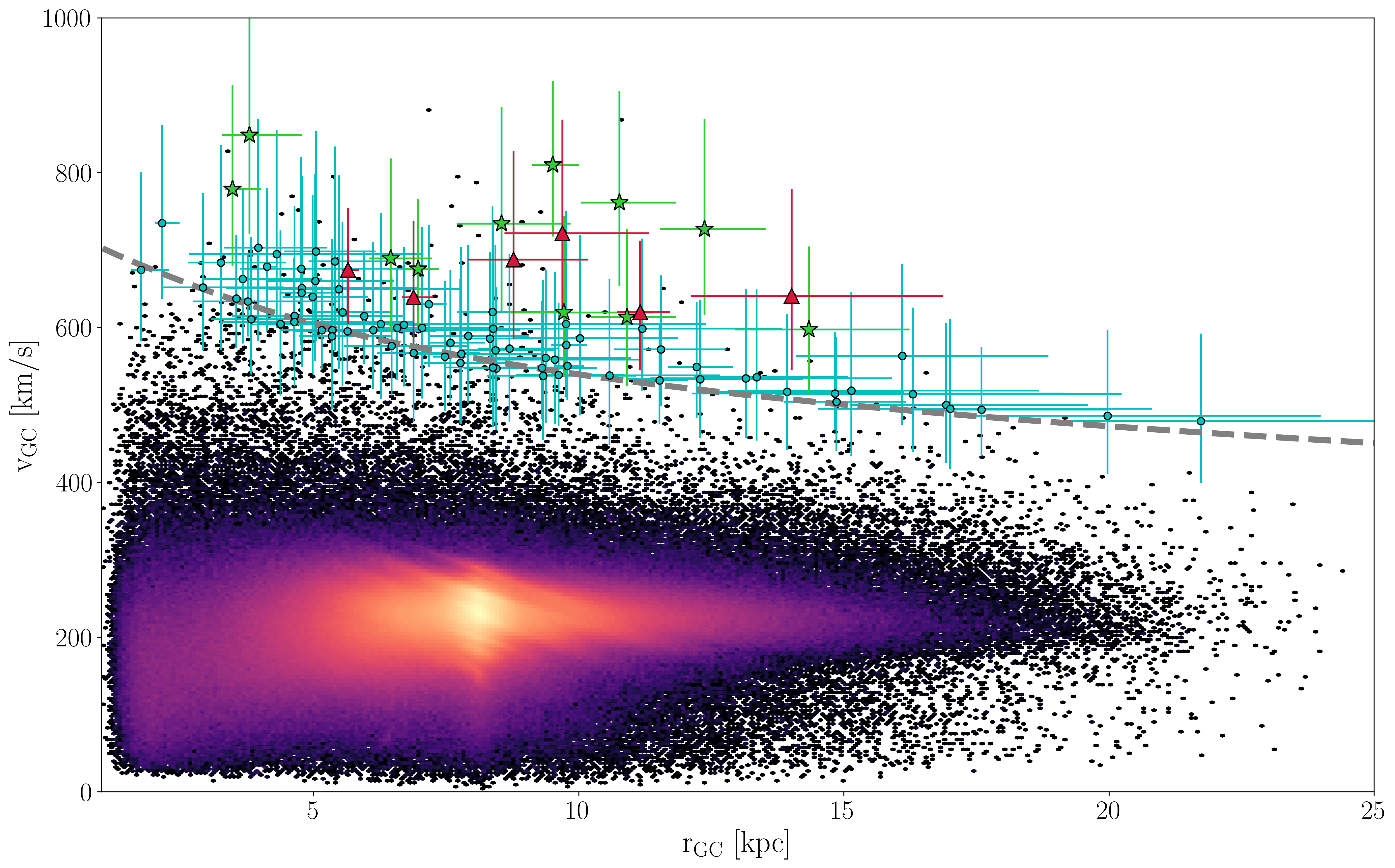}
	\caption{Total velocity in the Galactocentric frame as a function of distance from the Galactic Centre for the $\sim 7$ million stars with a radial velocity from {\Gaia} DR2, and with positive and precise parallaxes from {\Gaia} EDR3. Colour is proportional to the logarithm of the density of sources in each bin. The grey dashed line marks the escape speed from the Galaxy, computed using the \textsc{gala} potential \textsc{MilkyWayPotential} \citep[][refer to Section \ref{sec:orbits}]{gala}. Light blue points with errorbars represent our clean sample of 94 high velocity stars with $P_\mathrm{ub} > 0.5$. Green stars mark the subset of 11 Galactic stars with $P_\mathrm{ub}>0.8$ and $P_\mathrm{MW} > 0.5$, and red triangles the 6 extragalactic stars with $P_\mathrm{ub}>0.8$ and $P_\mathrm{MW} < 0.5$ (see Section \ref{sec:high_vel}).}
	\label{fig:vGC_rGC}
\end{figure*}

\begin{figure}
	\centering
 	\includegraphics[width=\columnwidth]{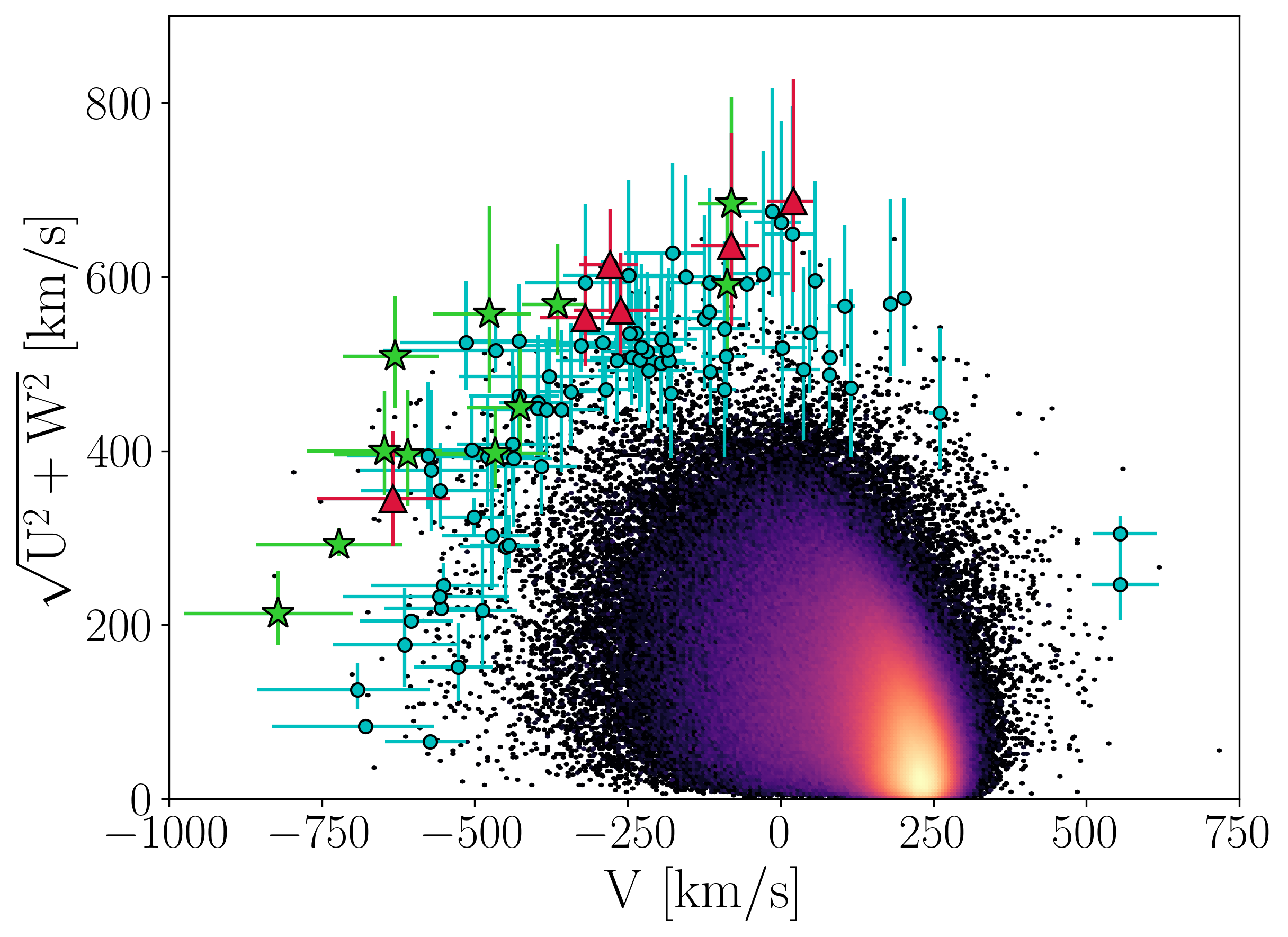}
	\caption{Toomre diagram for all the stars analyzed in this work. Colours and markers are the same as in Fig. \ref{fig:vGC_rGC}.}
	\label{fig:Toomre}
\end{figure}

\begin{figure*}
	\centering
 	\includegraphics[width=\textwidth]{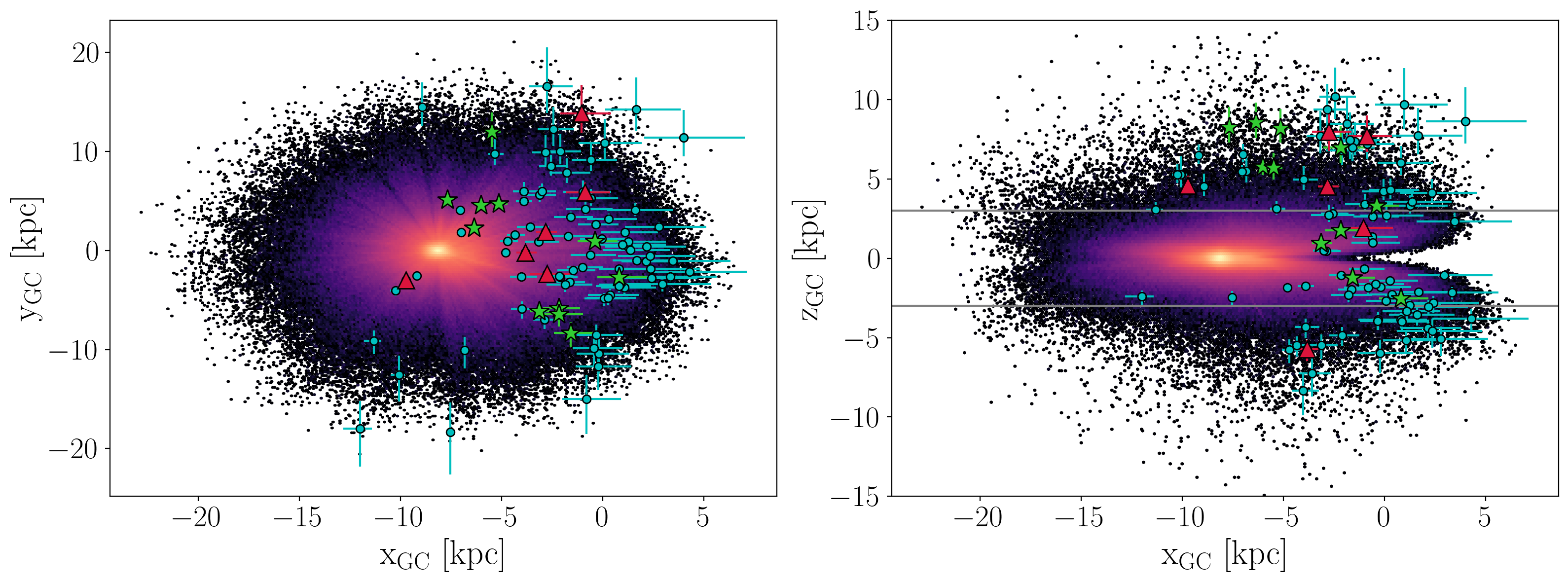}
	\caption{Distribution of stars on the Galactic plane (left panel) and in the $(x_\mathrm{GC}, z_\mathrm{GC})$ plane (right panel). Horizontal grey lines in the right panel mark the edges of the thick disk at $\pm 3$ kpc \citep{carollo+10}. The coordinates of the Sun are $(x_\odot, y_\odot, z_\odot) = (-8.211, 0, 0.025)$ kpc. Colours and markers are the same as in Fig. \ref{fig:vGC_rGC}.}
	\label{fig:yGC_xGC}
\end{figure*}

Following the methodology described in Section \ref{sec:method}, we can now study the three-dimensional spatial and velocity distributions of our sample of stars. We publish the resulting catalogue of positions and velocities as a single Flexible Image Transport System (FITS) file, publicly available \href{https://sites.google.com/view/tmarchetti/research}{here}. The list of the catalogue columns is provided in Appendix \ref{appendix:catalogue}. In the following, we will report and plot our results in terms of the median of the distribution, with lower and upper uncertainties computed using the $16$th and $84$th percentiles, respectively.

Fig. \ref{fig:vGC_hist} shows the histogram of the median total velocities in the Galactocentric rest-frame for all the stars considered in our analysis with a black line. Comparing this to the findings of \citetalias{Marchetti+19} (see their Figure 1), we note how the high velocity tail of the distribution extends to lower values. This is due to the fact that stars with extremely high velocities ($\gtrsim 1000$ {\kms}) are likely due to be artifacts resulting from noisy parallaxes, which we do not consider in this work.

In Fig. \ref{fig:vGC_rGC} we show the total velocity in the Galactocentric frame $v_\mathrm{GC}$ as a function of distance from the Galactic Centre $r_\mathrm{GC}$. The overdensity of sources at $r_\mathrm{GC} \sim 8$ kpc and $v_\mathrm{GC} \sim 250$ {\kms} corresponds to nearby bright disk stars, and we see how the density decreases with increasing velocities. The grey dashed line shows the escape velocity as a function of distance computed using the potential model described in Section \ref{sec:orbits}. In Fig. \ref{fig:Toomre} we show the Toomre diagram for the stars in our sample. This diagram is useful to distinguish between different populations of stars (thin disk, thick disk, and halo), based on their kinematics \citep[e.g.][]{venn+14}. On the $x$ axis we plot $V$, the Galactocentric component of the velocity along the direction of Galactic rotation, while on the $y$ axis the component orthogonal to $V$. We see that the majority of sources move on rotation supported, disk-like orbits centered on the value of the Local Standard of Rest \citep[LSR,][]{gaiadr2disk}. An extended and more diffuse population of stars with halo-like kinematics is visible centered at $V=0$, with a larger spread in $\sqrt{U^2 + W^2}$ \citep[e.g.][]{Bonaca+17, Koppelman+18, Yan+20}.

In Fig. \ref{fig:yGC_xGC} we plot the spatial distribution of the stars in our sample, with the left panel showing the distribution in Cartesian Galactocentric coordinates $(x_\mathrm{GC}, y_\mathrm{GC})$, and the right panel the distribution in $(x_\mathrm{GC}, z_\mathrm{GC})$. As expected, both plots show that the peak of the distribution is at the Sun's position at $-8.122$ kpc on the $x$ axis, and the majority of stars lie within a few kpc of the Galactic plane. Several features related to the presence of extinction are visible especially at low Galactic latitudes in the direction of the Galactic Centre. Comparing these figures to Figures 4 and 5 in \citetalias{Marchetti+19}, we notice how the combined effect of the increasing precision of {\Gaia} EDR3 parallaxes and the removal of star with noisy parallaxes results in a decrease in the number of sources at high Galactic latitudes and at large distances in the direction of (and beyond) the Galactic Centre.

Before discussing the population of stars with total velocities in excess of the local escape speed shown in Fig. \ref{fig:vGC_rGC}, in the following Section we will apply several quality cuts to the {\Gaia} astrometry and radial velocities, to select a clean sample of high velocity candidates.

\subsection{Selecting a clean sample of high velocity stars}
\label{sec:cuts}

\begin{table}
\caption{Summary of the cuts employed to select a clean subset of high velocity stars. The column $N_\mathrm{stars}$ denotes the number of stars in {\Gaia} EDR3 surviving each cut. The bulk of the analysis of this paper relies on the sample of 6969738 stars with precise astrometry from {\Gaia} EDR3 and radial velocities from {\Gaia} DR2.}           
\label{tab:cuts}      
\centering                          
\begin{tabular}{l r}        
\hline               
Cut & $N_\mathrm{stars}$ \\    
\hline
   Number of stars in {\Gaia} EDR3 & 1 811 709 771 \\
   $v_\mathrm{rad}$ from {\Gaia} DR2  & 7 209 831 \\
   parallax and proper motions from {\Gaia} EDR3 & 7 180 466 \\
   $\varpi > 0$, $\sigma_\varpi / \varpi < 20 \%$ & 6 969 738 \\
   \textsc{ruwe} $<1.4$ & 6 061 394 \\      
   \textsc{dr2\_rv\_nb\_transits} $\geq 4$ & 5 208 323 \\
   $\sigma_{v_\mathrm{GC}} / v_\mathrm{GC} < 30 \%$ & 5 191 458 \\
   $P_\mathrm{ub} > 50\%$ & 94 \\
   $P_\mathrm{ub} > 80\%$ & 17 \\
\hline                                   
\end{tabular}
\end{table}

\begin{figure}
	\centering
 	\includegraphics[width=\columnwidth]{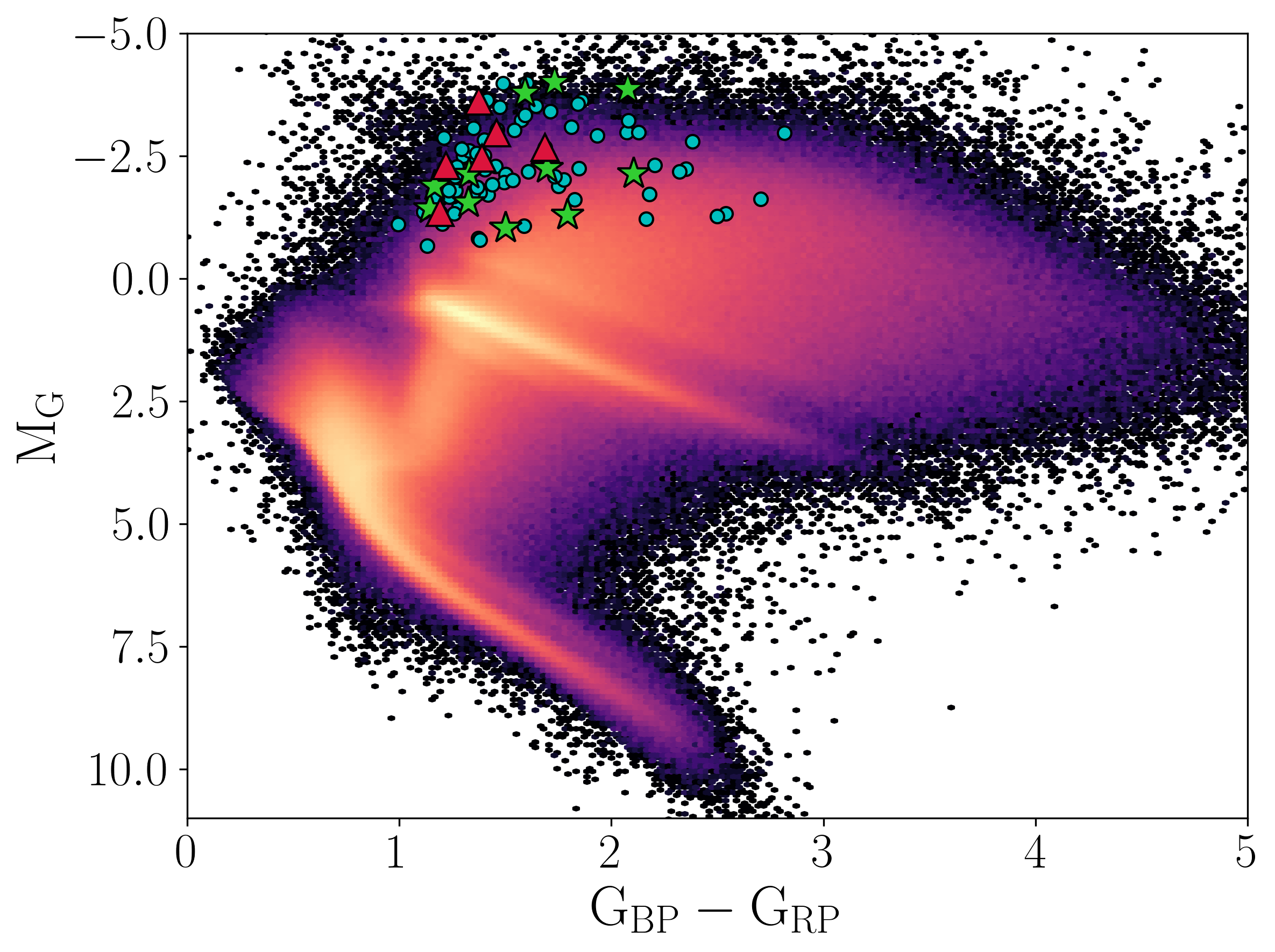}
	\caption{Colour-magnitude diagram for the stars considered in this work. Colours and markers are the same as in Fig. \ref{fig:vGC_rGC}.}
	\label{fig:HR}
\end{figure}

A selection of clean, well-behaved sources is essential when focusing on a subset of stars in a catalogues with million of entries. This is particularly true when searching for velocity outliers, since a wrong parallax, proper motion and/or radial velocity can easily result into a wrong high velocity object. We choose to apply the following quality cuts to our sample of stars, to remove possible outliers or instrumental artifacts:
\begin{enumerate}
    \item \textsc{ruwe} $< 1.4$;
    \item \textsc{dr2\_rv\_nb\_transits} $\geq 4$.
\end{enumerate}
The Renormalised Unit Weight Error (\textsc{ruwe}) is the recommended {\Gaia} EDR3 goodness-of-fit statistic to select astrometrically well-behaved sources \citep{Lindegren+20a}, and it is provided as a part of the {\Gaia} EDR3 main catalogue. This is the magnitude and colour renormalised version of the Unit Weight Error (UWE), which is defined as the square root of the reduced astrometric chi-squared statistic\footnote{see the {\Gaia} technical note GAIA-C3-TN-LU-LL-124-01.}. The distribution of this quantity should peak at $1$ for those sources for which the astrometric model (assuming single stars moving with uniform velocity) is a good fit to the observations. Possible sources showing an excess of \textsc{ruwe} include extended objects, variable stars, and multiple systems, for which the centre of light and the centre of mass do not coincide \citep[see][]{Belokurov+20, Penoyre+20}. \citet{Lindegren18} shows how the cut \textsc{ruwe} $<1.4$ selects the $70\%$ astrometrically best sources, which justifies our selection above.

Our further cut on the number of transits used to compute the radial velocity in {\Gaia} DR2 ensures that each star is observed a sufficient number of times for the radial velocity to be representative of the velocity of the system. In contrast with \citetalias{Marchetti+19}, here we follow \citet{Boubert+19}, that shows that a value $\geq 4$ retains a larger number of stars with a reliable radial velocity compared to the commonly used cut \textsc{dr2\_rv\_nb\_transits} $>5$. This cut also helps removing stars with blended spectra, which might have wrong radial velocities due to the presence of a nearby bright star \citep{Boubert+19}.

By applying the quality cuts described above, we are left with a total of 5208323 stars ($\sim 72 \%$ of the original sample) with reliable astrometry from {\Gaia} EDR3 and radial velocities from {\Gaia} DR2. We further select stars with relative errors in total velocities $\sigma_{v_\mathrm{GC}} / v_\mathrm{GC} < 30 \%$, where we compute $\sigma_{v_\mathrm{GC}}$ by summing in quadrature the lower and upper uncertainties on the total velocity of each star. 5191458 stars survive this additional cut. The velocity distribution of these stars is shown in purple in Fig. \ref{fig:vGC_hist}, where we can see that the cuts above exclude some of the fastest objects, which are likely to be instrumental artifacts, while retaining a few stars with $v_\mathrm{GC} > 700$ {\kms}. As a final cut, since we are interested in the fastest objects with high probabilities of being unbound from the Galaxy, we select sources with $P_\mathrm{ub} > 50\%$, for a total of 94 stars. 17 of these stars have $P_\mathrm{ub} > 80 \%$, and will be discussed in more details in Section \ref{sec:high_vel}. Our cuts are summarized in Table \ref{tab:cuts}, showing the number of stars surviving each cut.

Our clean sample of high velocity stars with $P_\mathrm{ub} > 0.5$ is marked in light blue in Fig. \ref{fig:vGC_hist} and following figures, where we also plot the uncertainties associated to this subset of stars. Looking at Fig. \ref{fig:vGC_rGC}, we see the presence of a few unbound objects with extremely high and precise total velocities that do not survive the quality cuts detailed above. These stars are likely not to be true high velocity stars, but "normal" stars with either a not reliable astrometry (\textsc{ruwe} $>1.4$) or a possibly wrong radial velocity (\textsc{dr2\_rv\_nb\_transits} $< 4$), which mimics a high velocity star. If we look at the sky distribution of the clean high velocity stars identified in this work, shown in Fig. \ref{fig:yGC_xGC}, we see that high velocity stars are observed mostly at high Galactic latitudes, in the direction of the Galactic Centre.

Fig. \ref{fig:HR} presents the observational Hertzsprung–Russell (HR) diagram for all the $\sim 7$ million sources considered in this work, showing the absolute magnitude in the {\Gaia} EDR3 $G$ band (computed assuming the median distance derived in Section \ref{sec:method}) as a function of the colour in the {\Gaia} EDR3 blue pass $G_\mathrm{BP}$ and red pass $G_\mathrm{RP}$. Most of the high velocity stars are identified to be giants, consistently with the {\Gaia} DR2 RVS results presented in \citetalias{Marchetti+19}, with the discussion in \citet{hattori+18b}, and with the results from high resolution spectroscopy presented in \citet{hawkins+18}. Note that we do not correct magnitudes and colours for extinction. The absence of early type high velocity stars is not surprising, since {\Gaia} DR2 provides radial velocities for stars with effective temperatures in the range $[3550, 6900]$ K \citep{Katz+19}.

\section{Orbit integration}
\label{sec:orbits}

\begin{table}
\centering
\caption{Characteristic scale parameters used in the \textsc{gala} potential \textsc{MilkyWayPotential} \citep{bovy15, gala}.}
\label{tab:pot_params}
\begin{tabular}{l|l}
	\hline
    Component & Parameters \\
    \hline
    Bulge & $M_b = 5.00 \cdot 10^9$ M$_\odot$ \\
     & $r_b = 1.00$ kpc \\[0.1cm]
    Nucleus & $M_n = 1.71 \cdot 10^9$ M$_\odot$ \\
     & $r_n = 0.07$ kpc \\[0.1cm]
    Disk & $M_d = 6.80 \cdot 10^{10}$ M$_\odot$ \\
     & $a_d = 3.00$ kpc \\[0.1cm]
     & $b_d = 0.28$ kpc \\[0.1cm]
    Halo & $M_h = 5.40 \cdot 10^{11}$ M$_\odot$ \\
     & $r_s = 15.62$ kpc \\
    \hline
\end{tabular}
\end{table}

In this Section, we perform orbital integration for our sample of 94 clean high velocity stars with probabilities higher than $50 \%$ of being unbound from the Milky Way, using the \textsc{python} package \textsc{gala} \citep{gala}. We make use of the \textsc{gala} potential \textsc{MilkyWayPotential}, a mass model of the Milky Way consisting of four different components. The bulge and the nucleus are modelled as a Hernquist spheroid \citep{hernquist90}:
\begin{equation}
\label{eq:HernquistBulge}
\phi_i(r_\mathrm{GC}) = -\frac{G M_i}{r_\mathrm{GC} + r_i},
\end{equation}
with $i=b, n$ for the bulge and the nucleus, respectively. The stellar disk is modelled as a Miyamoto \& Nagai axisymmetric disk \citep{M&N75} in Cylindrical coordinates $(R_\mathrm{GC}, z_\mathrm{GC})$:
\begin{equation}
\label{eq:MNdisk}
\phi_d(R_\mathrm{GC}, z_\mathrm{GC}) = -\frac{G M_d}{\sqrt{R_\mathrm{GC}^2 + \Bigl(a_d + \sqrt{z_\mathrm{GC}^2 + b_d^2}\Bigr)^2}},
\end{equation}
and the dark matter halo follows the spherically symmetric Navarro-Frenk-White \citep[NFW,][]{nfw96} profile:
\begin{equation}
\label{eq:NFW}
\phi_h(r_\mathrm{GC}) = -\frac{G M_h}{r_\mathrm{GC}} \ln \Bigl(1 + \frac{r_\mathrm{GC}}{r_s}\Bigr).
\end{equation}
The characteristic parameters of each component are the best fit parameters to mass measurements of the Milky Way out to $\sim 150$ kpc \citep{bovy15}, and they are reported in Table \ref{tab:pot_params}.

We integrate the orbit of each star back in time for a total time of 1 Gyr, with a time step of $0.1$ Myr. Following the same method outlined in Section \ref{sec:method}, we draw 5000 MC realizations of the orbit of each star starting from the MC samples of the observables, and we then derive the eccentricity, the energy and the angular momentum.

\begin{figure}
	\centering
 	\includegraphics[width=\columnwidth]{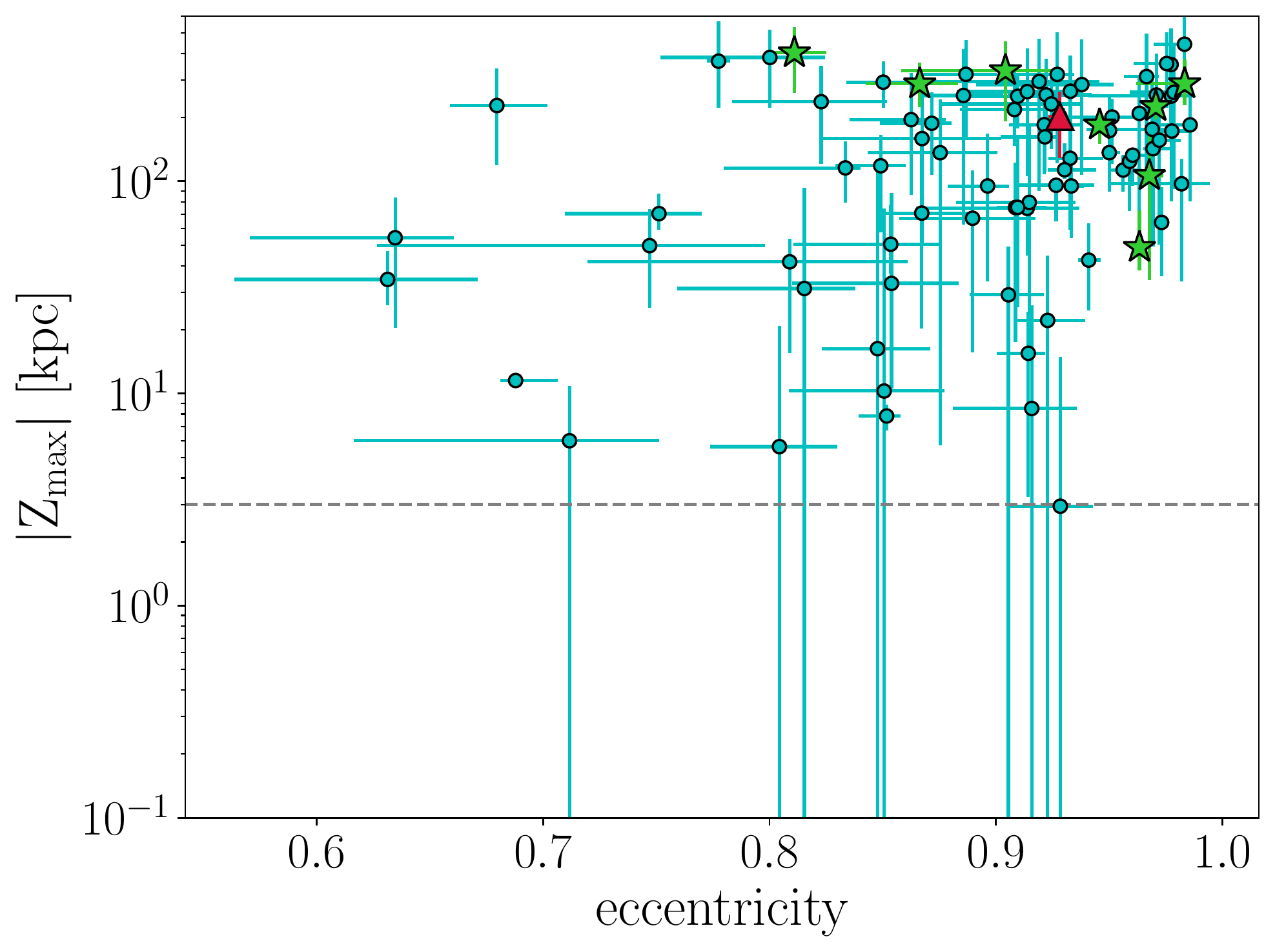}
	\caption{Maximum distance from the Galactic plane $|Z_\mathrm{max}|$ as a function of eccentricity for the clean sample of 94 high velocity stars with $P_\mathrm{ub} > 0.5$. 8 stars for which the eccentricity could not be determined from their orbits are not shown in this plot. Green stars (red triangles) mark the sample of Galactic (extragalactic) stars.}
	\label{fig:Zmax_ecc}
\end{figure}

In Fig. \ref{fig:Zmax_ecc} we plot the absolute value of the maximum distance from the Galactic plane $|Z_\mathrm{max}|$ as a function of the eccentricity of the orbit. This plot provides information on the shape of the orbit and of the vertical oscillations, and it has been used to identify stars moving on similar orbits \citep[e.g.][]{boeche+13,hawkins+15}. We note that $|Z_\mathrm{max}|$ is systematically above 3 kpc (horizontal dashed line) for the majority of our sources, corresponding to the edge of the thick disk \citep{carollo+10}. As expected, all of our candidates have very large eccentricities, compatible with the population of fast halo stars, and they can reach very large distances from the Galactic disk during their orbits, with values of $|Z_\mathrm{max}|$ up to several hundreds of kpc. We note that, since these stars are unbound in the majority of the MC realizations, the value of $|Z_\mathrm{max}|$ depends on the total integration time, which we arbitrarily set to 1 Gyr.

\begin{figure}
	\centering
 	\includegraphics[width=\columnwidth]{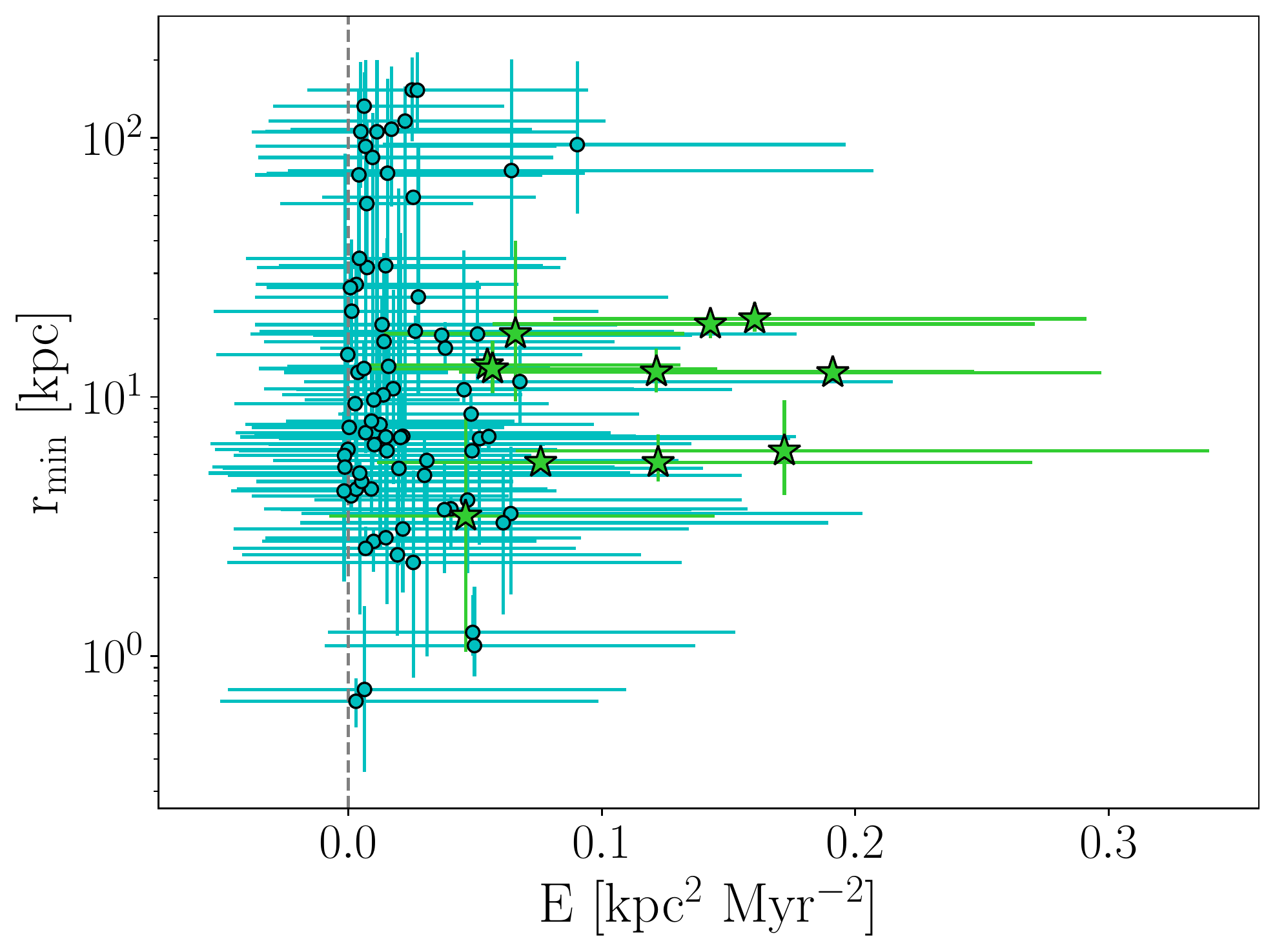}
	\caption{Minimum crossing radius $r_\mathrm{min}$ as a function of orbital energy per unit mass $E$ for the same stars plotted in Fig. \ref{fig:Zmax_ecc}. The vertical dashed line separates bound ($E<0$) from unbound ($E>0$) orbits. Colours and markers are the same as in Fig. \ref{fig:Zmax_ecc}. Extragalactic stars that never crossed the Galactic disk in the past 1 Gyr are not shown in this plot, since for these objects it is not possible to determine $r_\mathrm{min}$.}
	\label{fig:rmin_E}
\end{figure}

In our search for HVSs, we want to test whether our best candidates are consistent with coming from the very inner region of the Galaxy. For each orbit, we thus keep track of each Galactic disk crossing (Galactic latitude $b=0$), to identify the possible ejection location. Following \citetalias{Marchetti+19}, we define the crossing radius $r_\mathrm{c}$ as the distance from the Galactic Centre attained by the star during one disk crossing:
\begin{equation}
    \label{eq:rmin}
    r_\mathrm{c} \equiv \sqrt{x_\mathrm{c}^2 + y_\mathrm{c}^2} \ ,
\end{equation}
where $x_\mathrm{c}$ and $y_\mathrm{c}$ are, respectively, the $x$ and $y$ Cartesian Galactocentric coordinates at the moment of the disk crossing. Note that since these stars have high probabilities of being unbound from the Galaxy, there is a maximum of one crossing of the Galactic plane per orbit in the majority of the MC realizations. In the case of bound realizations with multiple disk crossings, we define $r_\mathrm{min}$ as the minimum value of $r_\mathrm{c}$. In Fig. \ref{fig:rmin_E} we plot the minimum crossing radius $r_\mathrm{min}$ as a function of the orbital energy per unit mass $E$. We see how our stars travel on unbound orbits $(E>0)$, and typical values of $r_\mathrm{min}$ range from a few kpc to more than $100$ kpc. Five stars are consistent with having $r_\mathrm{min} < 1$ kpc within the uncertainties, so the Galactic Centre cannot be excluded as the ejection location for these sources.

\section{High velocity star candidates}
\label{sec:high_vel}

In this Section we discuss our most promising candidates, the 17 stars with $P_\mathrm{ub} > 0.8$. Their {\Gaia} EDR3 identifiers, the observed properties from {\Gaia} EDR3, and a few quantities derived in this work are listed in Table \ref{tab:candidates}. Median total velocities range from $\sim 600$ {\kms} to more than $800$ {\kms}, and there are 12 stars with $P_\mathrm{ub} \geq 0.9$.

\begin{landscape}
\begin{table}
\centering
\caption{Observed and derived properties for the clean sample of 17 high velocity stars with a probability $P_\mathrm{ub} > 0.8$ of being unbound from the Milky Way. Quantities derived from the orbital integration are reported in Table \ref{tab:candidates_orb}. Stars are ordered by decreasing values of $P_\mathrm{ub}$.}
\label{tab:candidates}

\begin{threeparttable}
\setlength{\tabcolsep}{4pt}

\begin{tabular}{lcccccccccccc}

\hline
{\Gaia} EDR3 ID & (RA, Dec.) & ($l$, $b$) & $\varpi$ & $\mu_{\alpha*}$ & $\mu_\delta$ & $v_\mathrm{rad}$ & $G$ & $d$ & $r_\mathrm{GC}$ & $v_\mathrm{GC}$ & $P_\mathrm{MW}$ & $P_\mathrm{ub}$ \\
 & ($^\circ$) & ($^\circ$) & (mas) & (mas yr$^{-1}$) & (mas yr$^{-1}$) & (km s$^{-1}$) & (mag) & (pc) & (pc) & (\kms) &  &  \\
\hline

\textbf{Galactic} \\[0.15cm]

1383279090527227264 & ($240.337$, $41.167$) & ($65.46$, $48.85$) & $0.131 \pm 0.013$ & $-25.726 \pm 0.013$ & $-9.655 \pm 0.016$ & $-180.902 \pm 2.421$ & $13.0$ & $7644^{+834}_{-712}$ & $9511^{+499}_{-387}$ & $810^{+109}_{-93}$ & $1.0$ & $1.0$ \\[0.15cm] 

1375165725506487424 & ($231.849$, $36.034$) & ($57.9$, $55.86$) & $0.101 \pm 0.013$ & $-15.14 \pm 0.011$ & $-14.147 \pm 0.015$ & $-84.773 \pm 0.435$ & $11.19$ & $9895^{+1466}_{-1093}$ & $10767^{+1065}_{-728}$ & $761^{+144}_{-107}$ & $0.92$ & $0.99$ \\[0.15cm] 

1591615309672292224 & ($222.64$, $49.359$) & ($85.17$, $58.12$) & $0.103 \pm 0.014$ & $-15.377 \pm 0.013$ & $-13.192 \pm 0.018$ & $-123.121 \pm 0.403$ & $11.08$ & $9706^{+1485}_{-1156}$ & $12376^{+1156}_{-848}$ & $727^{+142}_{-111}$ & $0.97$ & $0.98$ \\[0.15cm] 

6703116567945367808 & ($273.431$, $-50.811$) & ($343.16$, $-15.15$) & $0.104 \pm 0.014$ & $-7.239 \pm 0.013$ & $-22.238 \pm 0.012$ & $209.079 \pm 2.936$ & $13.38$ & $9669^{+1474}_{-1169}$ & $3789^{+998}_{-524}$ & $849^{+161}_{-127}$ & $1.0$ & $0.97$ \\[0.15cm]

4337459232822884992 & ($249.981$, $-10.467$) & ($6.8$, $23.1$) & $0.118 \pm 0.015$ & $-19.142 \pm 0.017$ & $-14.857 \pm 0.013$ & $-290.985 \pm 0.411$ & $12.5$ & $8454^{+1226}_{-924}$ & $3469^{+537}_{-148}$ & $779^{+134}_{-99}$ & $1.0$ & $0.96$ \\[0.15cm] 

5847216962695435392 & ($207.66$, $-68.85$) & ($308.26$, $-6.59$) & $0.094 \pm 0.014$ & $-15.919 \pm 0.012$ & $-7.314 \pm 0.016$ & $260.077 \pm 18.159$ & $14.11$ & $10650^{+1822}_{-1365}$ & $8543^{+1301}_{-835}$ & $734^{+151}_{-113}$ & $1.0$ & $0.94$ \\[0.15cm] 

6065230602133664000 & ($203.583$, $-55.6$) & ($309.01$, $6.77$) & $0.125 \pm 0.012$ & $-21.301 \pm 0.012$ & $-6.024 \pm 0.013$ & $103.959 \pm 1.088$ & $13.22$ & $7977^{+862}_{-718}$ & $6972^{+400}_{-267}$ & $675^{+90}_{-75}$ & $0.65$ & $0.92$ \\[0.15cm] 

6090995247640198016 & ($212.05$, $-49.143$) & ($315.62$, $11.83$) & $0.117 \pm 0.018$ & $-16.81 \pm 0.02$ & $8.515 \pm 0.02$ & $21.091 \pm 0.523$ & $12.4$ & $8502^{+1509}_{-1125}$ & $6452^{+788}_{-406}$ & $689^{+129}_{-95}$ & $1.0$ & $0.87$ \\[0.15cm] 

2123569205674089856 & ($275.497$, $49.44$) & ($77.63$, $24.87$) & $0.074 \pm 0.011$ & $-9.826 \pm 0.012$ & $3.583 \pm 0.015$ & $-87.772 \pm 0.543$ & $11.66$ & $13505^{+2227}_{-1717}$ & $14343^{+1899}_{-1393}$ & $597^{+107}_{-82}$ & $1.0$ & $0.86$ \\[0.15cm] 

6193551030782723072 & ($201.24$, $-23.677$) & ($312.76$, $38.55$) & $0.09 \pm 0.015$ & $-13.149 \pm 0.017$ & $-0.749 \pm 0.01$ & $412.704 \pm 2.601$ & $13.12$ & $11172^{+2247}_{-1573}$ & $9718^{+1698}_{-1039}$ & $619^{+125}_{-83}$ & $1.0$ & $0.82$ \\[0.15cm] 

1477675943342041472 & ($213.484$, $31.892$) & ($52.76$, $71.46$) & $0.111 \pm 0.014$ & $-17.034 \pm 0.015$ & $-8.538 \pm 0.014$ & $71.542 \pm 5.089$ & $12.9$ & $9019^{+1300}_{-1037}$ & $10914^{+926}_{-680}$ & $613^{+114}_{-89}$ & $1.0$ & $0.82$ \\[0.15cm]

\textbf{Extragalactic} \\[0.15cm]

1297316350890352000 & ($248.178$, $21.272$) & ($38.97$, $39.59$) & $0.083 \pm 0.013$ & $-6.321 \pm 0.011$ & $-14.085 \pm 0.013$ & $-335.072 \pm 1.086$ & $12.96$ & $12053^{+2091}_{-1556}$ & $9693^{+1638}_{-1090}$ & $721^{+147}_{-107}$ & $0.0$ & $0.96$ \\[0.15cm]

3644492029415356032 & ($211.82$, $-4.143$) & ($336.18$, $53.76$) & $0.101 \pm 0.017$ & $7.943 \pm 0.018$ & $-12.983 \pm 0.014$ & $76.86 \pm 1.14$ & $11.36$ & $9896^{+2046}_{-1530}$ & $8773^{+1409}_{-869}$ & $687^{+141}_{-102}$ & $0.0$ & $0.93$ \\[0.15cm] 

4450458649852400640 & ($242.697$, $7.16$) & ($19.52$, $38.78$) & $0.139 \pm 0.014$ & $3.357 \pm 0.013$ & $-23.11 \pm 0.011$ & $-119.88 \pm 1.25$ & $12.94$ & $7206^{+751}_{-657}$ & $5647^{+211}_{-107}$ & $674^{+80}_{-69}$ & $0.0$ & $0.92$ \\[0.15cm] 

3859294747025975552 & ($161.032$, $6.306$) & ($241.66$, $53.29$) & $0.176 \pm 0.021$ & $4.94 \pm 0.025$ & $-27.251 \pm 0.025$ & $-98.225 \pm 5.147$ & $10.8$ & $5690^{+763}_{-628}$ & $11161^{+558}_{-441}$ & $620^{+93}_{-74}$ & $0.0$ & $0.90$ \\[0.15cm] 

2038763839481341056 & ($290.711$, $29.843$) & ($62.97$, $6.98$) & $0.064 \pm 0.011$ & $-4.285 \pm 0.01$ & $-8.479 \pm 0.012$ & $-9.172 \pm 3.509$ & $13.3$ & $15663^{+3231}_{-2285}$ & $14019^{+2849}_{-1898}$ & $641^{+138}_{-96}$ & $0.0$ & $0.90$ \\[0.15cm] 

6571360298580575104 & ($331.235$, $-42.701$) & ($357.05$, $-52.95$) & $0.139 \pm 0.016$ & $-13.125 \pm 0.013$ & $-18.851 \pm 0.014$ & $-1.458 \pm 1.936$ & $11.98$ & $7188^{+937}_{-746}$ & $6883^{+367}_{-212}$ & $639^{+98}_{-78}$ & $0.0$ & $0.83$ \\[0.15cm]

\hline
\end{tabular}

\begin{tablenotes}

\item \emph{Note.} Distances and total velocities are quoted in terms of the median of the distribution, with uncertainties derived from the $16$th and $84$th percentiles.

\end{tablenotes}

\end{threeparttable}
\end{table}
\end{landscape}

In addition to the cuts discussed in Section \ref{sec:cuts}, we check that none of these stars is flagged as a \textsc{duplicated\_source} in {\Gaia} EDR3 (which might point to multiplicity or to cross-matching problems in crowded fields). In addition, we inspect the {\Gaia} EDR3 column \textsc{ipd\_gof\_harmonic\_amplitude}, which measures the amplitude of the variation of the reduced chi-square of the Image Parameters Determination (IPD) module as a function of the scan direction. A large value might be due to multiplicity of the source, which might in turn affect the astrometry and radial velocity determination. We check that this value is $< 0.05$ for all of our sources.

\noindent
Finally, we check the distribution of the {\Gaia} EDR3 column \textsc{visibility\_periods\_used}. A visibility period is defined as a group of observations separated by more then 4 days from the other groups, and a value greater then 8 is required internally for the full astrometric solution \citep{Lindegren+18, Lindegren18, Lindegren+20a}.
All of our sources have \textsc{visibility\_periods\_used} $\geq 13$, suggesting again a reliable astrometry. Furthermore, all the stars have \textsc{dr2\_rv\_nb\_transits} $\geq 5$. In addition, the {\Gaia} EDR3 \textsc{source\_id} is equal to the {\Gaia} DR2 \textsc{source\_id} for all of the sources.

In \citetalias{Marchetti+19} we found that more than half of the stars with probabilities $\geq 80\%$ from the Galaxy have orbits that never intersected the Galactic disk in the past, therefore they are not consistent with being ejected from a known Galactic star forming region. These stars were therefore labelled as \emph{extragalactic}. Here we follow the same approach as in \citetalias{Marchetti+19}, and we divide our sample of 17 clean high velocity stars with $P_\mathrm{ub}>80\%$ in Galactic and extragalactic stars. We define the probability of coming from the Milky Way disk $P_\mathrm{MW}$ as the fraction of MC realizations resulting into the orbit of the star intersecting the Galactic disk, which translates into the following condition on the minimum crossing radius: $r_\mathrm{min} < 25$ kpc \citep{xu+15}. A high value of $P_\mathrm{MW}$ is a necessary condition for both HVS and hyper-runaway star candidates. We define as Galactic stars those stars with $P_\mathrm{MW} > 0.5$, and as extragalactic stars those sources satisfying $P_\mathrm{MW}<0.5$. The best unbound candidates discovered in \citetalias{Marchetti+19} are revisited using the new {\Gaia} EDR3 astrometry in Appendix \ref{appendix:highV_DR2}.

\begin{table*}
	
	\caption{Galactocentric cylindrical coordinates $(R_\mathrm{GC}, z_\mathrm{GC})$,  energy per unit mass $E$, minimum crossing radius $r_\mathrm{min}$, ejection velocity $v_\mathrm{ej}$, and flight time $t_\mathrm{F}$ for the clean sample of 17 stars with $P_\mathrm{ub} > 0.8$. Stars are ordered as in Table \ref{tab:candidates}.}
	
	\label{tab:candidates_orb}
	
	\begin{threeparttable}
	\setlength{\tabcolsep}{12pt}
		
		\begin{tabular}{lcccccc}
		
		\hline
        {\Gaia} EDR3 ID & $R_\mathrm{GC}$ & $z_\mathrm{GC}$ & $E$ & $r_\mathrm{min}$ & $v_\mathrm{ej}$ & $t_\mathrm{F}$ \\
         & (pc) & (pc) & (kpc$^2$ Myr$^{-2}$) & (pc) & (km s$^{-1}$) & (Myr) \\
\hline

\textbf{Galactic} \\[0.15cm]

1383279090527227264 & $7558^{+138}_{-88}$ & $5774^{+627}_{-535}$ & $0.191^{+0.106}_{-0.074}$ & $12412^{+1163}_{-908}$ & $796^{+113}_{-89}$ & $10.9^{+0.3}_{-0.2}$\\[0.15cm] 

1375165725506487424 & $6972^{+191}_{-81}$ & $8205^{+1212}_{-904}$ & $0.16^{+0.131}_{-0.079}$ & $20020^{+3117}_{-2204}$ & $728^{+148}_{-104}$ & $21.0^{+0.2}_{-0.2}$\\[0.15cm]

1591615309672292224 & $9211^{+400}_{-276}$ & $8265^{+1261}_{-981}$ & $0.143^{+0.128}_{-0.086}$ & $19106^{+2819}_{-2214}$ & $708^{+148}_{-118}$ & $19.7^{+0.5}_{-0.4}$\\[0.15cm] 

6703116567945367808 & $2821^{+973}_{-412}$ & $-2529^{+309}_{-389}$ & $0.172^{+0.167}_{-0.106}$ & $6188^{+3522}_{-2005}$ & $819^{+152}_{-104}$ & $10.4^{+1.8}_{-1.2}$\\[0.15cm] 

4337459232822884992 & $1198^{+504}_{-210}$ & $3317^{+478}_{-360}$ & $0.122^{+0.125}_{-0.078}$ & $12498^{+2840}_{-2086}$ & $708^{+141}_{-99}$ & $16.0^{+0.2}_{-0.2}$\\[0.15cm] 

5847216962695435392 & $8456^{+1284}_{-821}$ & $-1218^{+159}_{-213}$ & $0.122^{+0.148}_{-0.087}$ & $5582^{+1571}_{-868}$ & $766^{+150}_{-104}$ & $5.9^{+0.2}_{-0.2}$\\[0.15cm] 

6065230602133664000 & $6907^{+390}_{-258}$ & $949^{+100}_{-83}$ & $0.066^{+0.067}_{-0.052}$ & $17538^{+22501}_{-7955}$ & $596^{+33}_{-30}$ & $35.4^{+30.32}_{-10.4}$\\[0.15cm] 

6090995247640198016 & $6211^{+732}_{-359}$ & $1749^{+306}_{-228}$ & $0.076^{+0.102}_{-0.065}$ & $5593^{+776}_{-381}$ & $706^{+118}_{-88}$ & $3.1^{+0.0}_{-0.0}$\\[0.15cm] 

2123569205674089856 & $13163^{+1663}_{-1207}$ & $5697^{+935}_{-721}$ & $0.055^{+0.076}_{-0.05}$ & $13305^{+2065}_{-1497}$ & $606^{+100}_{-74}$ & $9.6^{+0.0}_{-0.1}$\\[0.15cm] 

6193551030782723072 & $6771^{+994}_{-493}$ & $6970^{+1397}_{-978}$ & $0.046^{+0.098}_{-0.054}$ & $3468^{+4863}_{-2429}$ & $702^{+63}_{-9}$ & $17.4^{+3.6}_{-2.2}$\\[0.15cm] 
 
1477675943342041472 & $6758^{+109}_{-116}$ & $8570^{+1232}_{-983}$ & $0.057^{+0.089}_{-0.055}$ & $12836^{+3625}_{-2507}$ & $612^{+109}_{-73}$ & $23.3^{+1.2}_{-1.0}$\\[0.15cm]

\textbf{Extragalactic} \\[0.15cm]

1297316350890352000 & $5907^{+959}_{-506}$ & $7685^{+1329}_{-989}$ & $0.124^{+0.132}_{-0.087}$ & - & - & - \\[0.15cm]

3644492029415356032 & $3648^{+465}_{-322}$ & $7990^{+1647}_{-1231}$ & $0.099^{+0.118}_{-0.07}$ & - & - & - \\[0.15cm] 

4450458649852400640 & $3382^{+330}_{-316}$ & $4522^{+469}_{-410}$ & $0.058^{+0.071}_{-0.046}$ & - & - & - \\[0.15cm] 

3859294747025975552 & $10173^{+328}_{-263}$ & $4591^{+612}_{-504}$ & $0.059^{+0.071}_{-0.047}$ & - & - & - \\[0.15cm] 

2038763839481341056 & $13889^{+2822}_{-1878}$ & $1906^{+388}_{-274}$ & $0.09^{+0.106}_{-0.077}$ & - & - & - \\[0.15cm] 

6571360298580575104 & $3821^{+445}_{-558}$ & $-5725^{+597}_{-750}$ & $0.049^{+0.079}_{-0.051}$ & - & - & - \\[0.15cm]

\hline
		
		\end{tabular}
		
		\begin{tablenotes}

\item \emph{Note.} Derived quantities are quoted in terms of the median of the distribution, with uncertainties derived from the $16$th and $84$th percentiles.

\end{tablenotes}
		
	\end{threeparttable}
	
\end{table*}

\begin{figure}
	\centering
 	\includegraphics[width=\columnwidth]{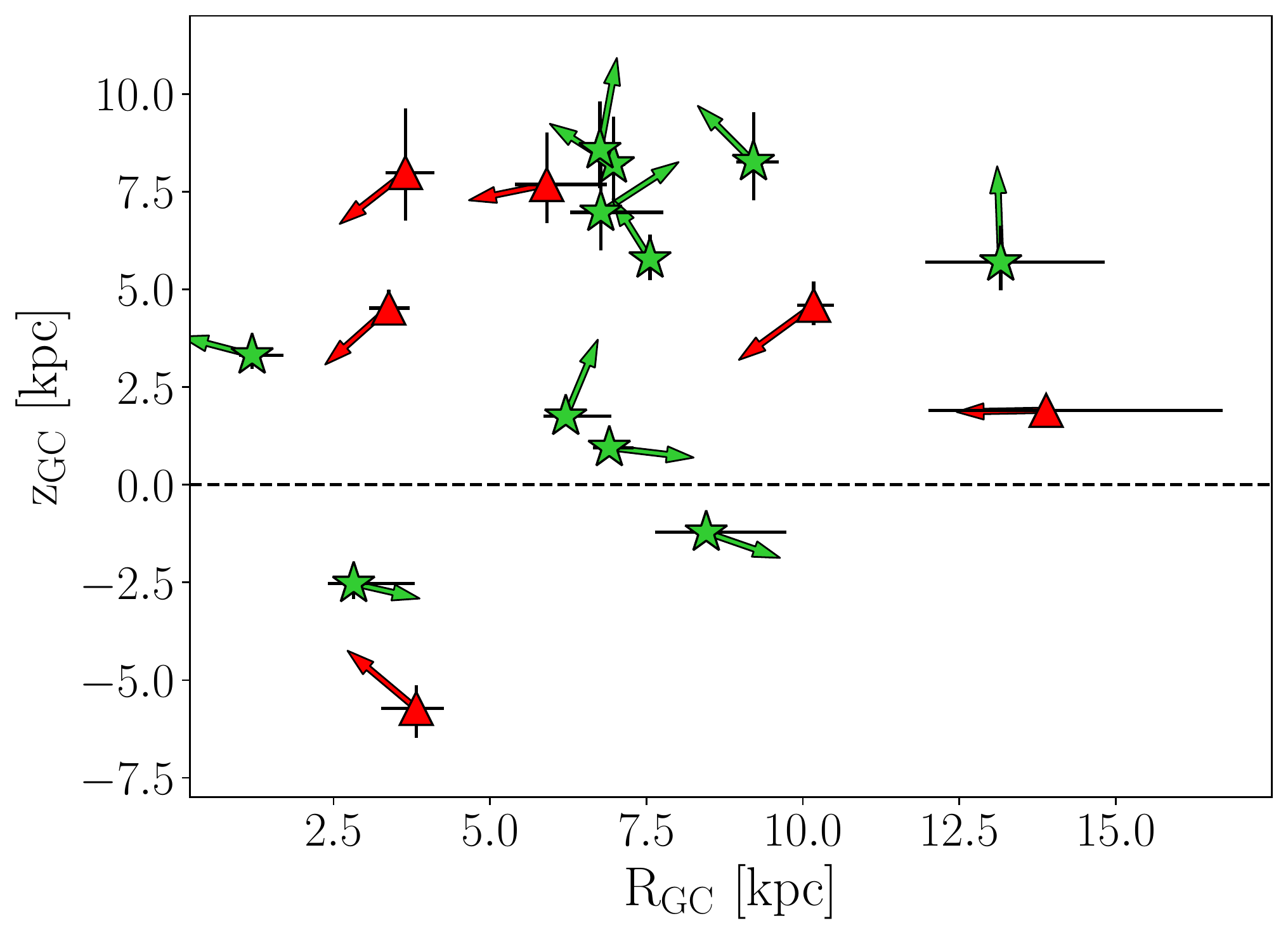}
	\caption{Distribution in Galactocentric cylindrical coordinates $(R_\mathrm{GC}, z_\mathrm{GC})$ of the clean sample of high velocity stars with $P_\mathrm{ub} > 0.8$. Green stars mark the Galactic stars, while red triangles mark the extragalactic sources. Arrows point in the direction of the total velocity vector $\mathbf{v_\mathrm{GC}}$, and their length is proportional to $v_\mathrm{GC}$. The horizontal dashed line at $z_\mathrm{GC}=0$ marks the position of the Galactic disk. The Sun is at $(R_\mathrm{GC}, z_\mathrm{GC}) = (8.122, 0.025)$ kpc.}
	\label{fig:arrows}
\end{figure}

Table \ref{tab:candidates_orb} shows Galactocentryc cylindrical coordinates, energy per unit mass $E$, the minimum crossing radius $r_\mathrm{min}$, the ejection velocity $v_\mathrm{ej}$, and the flight time $t_\mathrm{F}$ for the sample of stars with $P_\mathrm{ub} > 0.8$. We compute the ejection velocity as the total velocity of the star at the moment of crossing the Galactic disk, and the flight time as the time spent between the disk crossing and the current observation of the source. The minimum crossing radius, the ejection velocity and the flight time are defined only for the sample of Galactic stars. The position of these stars in Galactocentric cylindrical coordinates is shown in Fig. \ref{fig:arrows}, where the arrows point in the direction of the total velocity vector. From this plot, we can clearly see that Galactic stars point away from the Galactic disk (horizontal dashed line at $z_\mathrm{GC}=0$), while extragalactic stars are currently pointing away from an unknown star forming region. In the following, we will present and discuss the most interesting objects.

\subsection{Galactic stars}
\label{sec:galactic}

11 of the clean sample of 17 stars with $P_\mathrm{ub}>0.8$ have $P_\mathrm{MW}>0.5$, and are therefore labelled as Galactic stars. They are marked as green stars in Fig. \ref{fig:vGC_rGC} and following figures. These stars could be either HVS or hyper-runaway star candidates. Looking at the distribution of the minimum crossing radii, shown in Fig. \ref{fig:rmin_E}, we can see that the minimum crossing radius is around $3.5$ kpc, with typical values of several tens of kpc. We can therefore conclude that the Galactic Centre is excluded as the ejection location of these stars, and therefore there are no HVS candidates, consistently with predictions from simulations \citep{marchetti+18}. We will discuss this further in Section \ref{sec:discuss}.

Ejection velocities are in the range $\sim 600 - 820$ {\kms}, challenging for the current model predictions or hyper-runaway stars. We note that all ejection velocities are higher than the current upper limit of $\sim 450$ {\kms} for early type stars, but this is not surprising since our sources are likely to be giants (see Fig. \ref{fig:HR}). Flight times from the stellar disk are of the order of a few tens of Myr, consistent with the extreme observed velocities, and are much shorter then typical lifetimes of giant stars. All of the Galactic stars have probabilities $P_\mathrm{MW} > 0.9$ except for {\Gaia} EDR3 6065230602133664000, which has $P_\mathrm{MW} \sim 0.6$. This can be seen in Fig. \ref{fig:arrows}, where its velocity vector is almost parallel to (but slightly point towards) the Galactic disk.

{\Gaia} EDR3 1383279090527227264 was already identified as a Galactic hyper-runaway star in \citetalias{Marchetti+19}, \citet{Bromley+18} and \citet{Du+18}. With a total velocity of more than $800$ {\kms} and a probability of being unbound $P_\mathrm{ub}=1$, this star is our strongest hyper-runaway candidate\footnote{This is also confirmed when correcting for the parallax zero point, as reported in Section \ref{sec:zp}.}. The new distance derived from the {\Gaia} EDR3 parallax is $\sim 7.7$ kpc, lower than the $8.5$ kpc reported in \citetalias{Marchetti+19} and \citet{Bromley+18}. Its ejection velocity is estimated to be $\sim 800$ {\kms}, and it was ejected from the Galactic disk $\sim 11$ Myr ago. Another possible origin for this object is discussed in \citet{Bromley+18}, which suggest that this star could have originated from the LMC, having an orbit passing $\sim 15$ kpc from the LMC centre, $\sim 70$ Myr ago. We test this possibility using the new {\Gaia} EDR3 astrometry. We find that the minimum distance of its trajectory from the centre of the LMC is now $\sim 50$ kpc, which excludes the LMC as a possible origin for this star.

The other Galactic candidates are not reported in \citetalias{Marchetti+19}, \citet{Bromley+18}, \citet{hattori+18b} and \citet{Du+18}, so they are new hyper-runaway star candidates from {\Gaia} EDR3. 

\subsection{Extragalactic stars}
\label{sec:extragalactic}

The other 6 of the 17 stars with $P_\mathrm{ub}>0.8$ have $P_\mathrm{MW}<0.5$, and therefore an extragalactic origin is preferred for these sources. These stars are marked as red triangles in Fig. \ref{fig:vGC_rGC} and following figures. Looking at Fig. \ref{fig:arrows}, it is evident that these sources are pointing away from the stellar disk. Therefore, if confirmed, these objects need to have formed either in the stellar halo \citep[e.g.][]{Hambly+96} or to come from a disrupting satellite \citep[e.g.][]{abadi+09}. 

Total velocities of these stars range from $\sim 620$ {\kms} to $\sim 720$ {\kms}, and their distribution in Galactic coordinates in Fig. \ref{fig:yGC_xGC} shows that they are mostly observed at high Galactic latitudes. None of these stars is listed as an unbound candidate in \citetalias{Marchetti+19}, \citet{Bromley+18}, \citet{hattori+18b} or \citet{Du+18}.

{\Gaia} EDR3 2038763839481341056 is observed at a Galactic latitude $b \sim 7^\circ$ ($z_\mathrm{GC} \sim 1.9$ kpc), therefore, even it is classified as an extragalactic star, it is likely to have originated in the Galactic disk of the Milky Way. This is also confirmed by its trajectory, which runs parallel to the disk (see Fig. \ref{fig:arrows}).

\section{Parallax zero point}
\label{sec:zp}

\begin{figure}
	\centering
 	\includegraphics[width=\columnwidth]{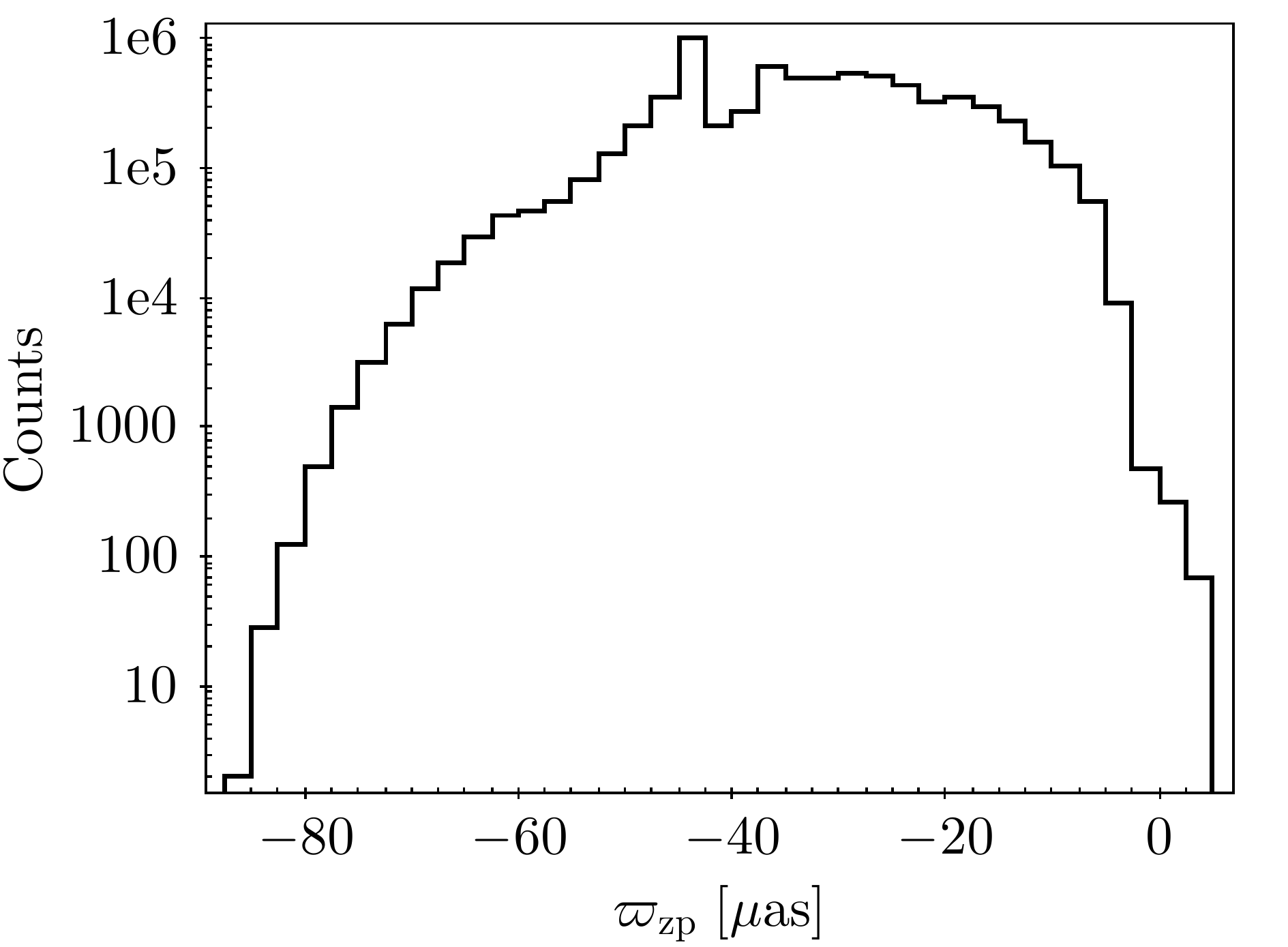}
	\caption{Histogram of the estimated parallax zero point $\varpi_\mathrm{zp}$ for the sample of 7180466 stars with {\Gaia} EDR3 astrometry and {\Gaia} DR2 radial velocities.}
	\label{fig:zp}
\end{figure}

In this Section we discuss the impact of assuming a zero point in {\Gaia} EDR3 parallaxes. The difference between the observed parallax of a star in {\Gaia} EDR3 and the true parallax of the source can be modelled as the sum of a random error $r$ with mean zero and variance given by the square root of the quoted parallax uncertainty, and a systematic error $s$ \citep{Lindegren+18, Arenou+18, Lindegren+20a}. The mean value of the systematic error $s$ is the parallax zero point $\varpi_\mathrm{zp}$. In {\Gaia} EDR3, the median value of $\varpi_\mathrm{zp}$ for quasars is $-17$ $\mu$as \citep{Lindegren+20b}. In principle, $\varpi_\mathrm{zp}$ is expected to be a complex function of the magnitude of the source, its colour, its position on the sky, and possibly on a number of other variables. Following \citet{Lindegren+20b}, this can be modelled as being dependent on the {\Gaia} EDR3 columns \textsc{phot\_g\_mean\_mag} (the magnitude of the star in the {\Gaia} EDR3 $G$ band), \textsc{ecl\_lat} (the ecliptic latitude of the star), \textsc{pseudocolour} (the pseudo-colour of the source estimated using the astrometry), \textsc{nu\_eff\_used\_in\_astrometry} (the effective wavenumber of the star used in the astrometric fit), and \textsc{astrometric\_params\_solved} (indicating which astrometric parameters were determined for the star). Estimating the zero point for each star in our sample, we find that typical values of $\varpi_\mathrm{zp}$ range from $-80$ $\mathrm{\mu}$as to $\sim 0$, as shown in Fig. \ref{fig:zp}. The median of the distribution is at $-33$ $\mu$as, and the mode is around $-43$ $\mu$as.

We subtract the estimated parallax zero point to the quoted {\Gaia} EDR3 parallax for all the 7180466 sources with full {\Gaia} EDR3 astrometry and radial velocities from {\Gaia} DR2. Then, we derive distances and total velocities for the sample of stars with $\sigma_\varpi/(\varpi-\varpi_\mathrm{zp})<0.2$, for a total of 7065053 sources. For the subset of 221406 sources with 6-parameter solutions ($\sim 3\%$), we draw 5000 MC samples from a multi-variate Gaussian distribution, with mean vector:
\begin{equation}
    \label{eq:mean_6p}
    \mathbf{m} = [\mu_{\alpha*}, \mu_\delta, \varpi, \hat{\nu}_\mathrm{eff}] \ ,
\end{equation}
where $\hat{\nu}_\mathrm{eff}$ is the pseudo-colour of the source. The covariance matrix is now expressed as a $4\times 4$ symmetric matrix, with diagonal elements $\Sigma_{ii} = \sigma_i^2$, and off-diagonal elements $\Sigma_{ij} = \sigma_i\sigma_j\rho(i,j)$, with $i, j$ being the astrometric parameters $\mu_{\alpha*}, \mu_\delta, \varpi, \hat{\nu}_\mathrm{eff}$. We then subtract to each parallax sample the zero point value estimated using the corresponding pseudo-colour sample, and we derive distances and total velocities as described in Section \ref{sec:method}. The approach used for the sources with 5-parameter solutions is equivalent to the one outlined in Section \ref{sec:method}.

By applying the same quality cuts discussed in Section \ref{sec:cuts}, we find a total of 12 stars with reliable {\Gaia} measurements and with $P_\mathrm{ub} > 0.5$. 3 of them have $P_\mathrm{ub}>0.8$. The observational and derived properties of the 12 stars with $P_\mathrm{ub}>0.5$ are presented in Table \ref{tab:candidates_zp}. Following the approach discussed in Section \ref{sec:high_vel}, 7 of these sources are classified as Galactic hyper-runaway star candidates, and 5 as extragalactic stars. $4$ Galactic sources ({\Gaia} EDR3 4337459232822884992, {\Gaia} EDR3 6090995247640198016, {\Gaia} EDR3 1383279090527227264, and {\Gaia} EDR3 1375165725506487424) and 1 extragalactic star ({\Gaia} EDR3 4182243409716233856) are in common with the candidates selected in Section \ref{sec:high_vel} and presented in Tab \ref{tab:candidates}. For these stars, the classification in Galactic and extragalactic sources does not depend on the inclusion of the parallax zero point. We note that velocities are lower than those reported in Table \ref{tab:candidates}, since the effect of the negative offset in parallax is to lower derived distances, and therefore total velocities.

\section{Discussion and conclusions}
\label{sec:discuss}

We have fully characterized the high velocity tail of the velocity distribution of $\sim 7$ million stars with precise astrometry from {\Gaia} EDR3 and radial velocities from {\Gaia} DR2. In Section \ref{sec:zp} we have investigated the impact of assuming a zero point correction to {\Gaia} EDR3 parallaxes, using the approach outlined in \citet{Lindegren+20b}. We have derived positions and velocities in the Galactocentric frame for all the stars, which are available in a FITS catalogue accessible \href{https://sites.google.com/view/tmarchetti/research}{here}. In Appendix \ref{appendix:catalogue} we detail the catalogue content. We have further selected stars with reliable astrometry and radial velocities, focusing on objects with a high probability of being unbound from our Galaxy. We obtained a clean sample of 94 high velocity giant stars with $P_\mathrm{ub} > 50\%$, with reliable astrometry and radial velocities. By back-propagating the orbits of the high velocity stars in the Galactic potential, we find that:
\begin{itemize}
    
    \item 17 of these stars have $P_\mathrm{ub} > 80\%$, and represent our best unbound star candidates. These stars are presented in Table \ref{tab:candidates}, and properties derived from their orbits are presented in Table \ref{tab:candidates_orb}. Fig. \ref{fig:arrows} shows the direction of the total velocity vector of these objects. If we include the effect of the parallax zero point, the number of stars with $P_\mathrm{ub}>50\%$ and $P_\mathrm{ub}>80\%$ reduces, respectively, to $12$ and $3$;
    \item 11 of the 17 stars are consistent with being ejected from the Galactic disk, and are therefore hyper-runaway star candidates. The analysis including the zero point correction to parallaxes identifies $7$ hyper-runaway star candidates with $P_\mathrm{ub}>0.5$, $3$ of those having $P_\mathrm{ub}>0.8$. Our best candidate is {\Gaia} EDR3 1383279090527227264, which was already identified in \citetalias{Marchetti+19}, \citet{Bromley+18} and \citet{Du+18}. This star has a total velocity of $\sim 820$ {\kms}, and we predict it to have been ejected with a velocity of $\sim 800$ {\kms} from the Galactic disk $\sim 11$ Myr ago;
    \item 6 stars with $P_\mathrm{ub}>0.8$, when traced back in time, do not intersect the Galactic plane, and therefore an extragalactic origin is preferred. The orbits of these stars are currently pointing towards the Galactic disk. We find $5$ extragalactic stars with $P_\mathrm{ub}>0.5$ once we correct parallaxes for the parallax zero point;
    \item Regardless of the choice of adopting or not the zero point correction to {\Gaia} EDR3 parallaxes, 0 unbound stars are consistent with coming from the Galactic Centre, so there are no HVS candidates.

\end{itemize}

\begin{landscape}
\begin{table}
\caption{Observed and derived properties for the clean sample of 12 high velocity stars with a probability $P_\mathrm{ub} > 0.5$ of being unbound from the Milky Way, computed including the estimate of the parallax zero point $\varpi_\mathrm{zp}$. Stars are ordered by decreasing values of $P_\mathrm{ub}$.}
\label{tab:candidates_zp}

\begin{threeparttable}
\setlength{\tabcolsep}{5pt}

\begin{tabular}{lccccccccccccc}

\hline
{\Gaia} EDR3 ID & (RA, Dec.) & ($l$, $b$) & $\varpi - \varpi_\mathrm{zp}$ & $\mu_{\alpha*}$ & $\mu_\delta$ & $v_\mathrm{rad}$ & $G$ & $d$ & $r_\mathrm{GC}$ & $v_\mathrm{GC}$ & $P_\mathrm{MW}$ & $P_\mathrm{ub}$ \\
 & ($^\circ$) & (mas) & (mas yr$^{-1}$) & (mas yr$^{-1}$) & (km s$^{-1}$) & (mag) & (pc) & (pc) & (\kms) &  &  \\
\hline

\textbf{Galactic} \\[0.15cm]

6014438731195297408 & ($233.205$, $-33.907$) & ($337.33$, $17.97$) & $0.118 \pm 0.017$ & $-15.224 \pm 0.018$ & $-7.848 \pm 0.014$ & $-253.194 \pm 8.513$ & $12.94$ & $10693^{+2307}_{-1609}$ & $5278^{+1766}_{-922}$ & $729^{+170}_{-112}$ & $1.0$ & $0.88$ \\[0.15cm] 

4337459232822884992 & ($249.981$, $-10.467$) & ($6.8$, $23.1$) & $0.14 \pm 0.015$ & $-19.142 \pm 0.017$ & $-14.857 \pm 0.013$ & $-290.985 \pm 0.411$ & $12.5$ & $7745^{+992}_{-752}$ & $3359^{+213}_{-48}$ & $702^{+107}_{-79}$ & $1.0$ & $0.86$ \\[0.15cm]

1383279090527227264 & ($240.337$, $41.167$) & ($65.46$, $48.85$) & $0.199 \pm 0.013$ & $-25.726 \pm 0.013$ & $-9.655 \pm 0.016$ & $-180.902 \pm 2.421$ & $13.0$ & $6071^{+521}_{-451}$ & $8710^{+243}_{-190}$ & $606^{+68}_{-59}$ & $1.0$ & $0.84$ \\[0.15cm] 

6090995247640198016 & ($212.05$, $-49.143$) & ($315.62$, $11.83$) & $0.139 \pm 0.018$ & $-16.81 \pm 0.02$ & $8.515 \pm 0.02$ & $21.091 \pm 0.523$ & $12.4$ & $7807^{+1239}_{-935}$ & $6180^{+528}_{-256}$ & $630^{+105}_{-78}$ & $1.0$ & $0.71$ \\[0.15cm] 

6901673112193397248 & ($309.967$, $-10.203$) & ($35.99$, $-28.68$) & $0.118 \pm 0.017$ & $6.933 \pm 0.018$ & $-10.591 \pm 0.013$ & $-455.382 \pm 2.084$ & $12.88$ & $10726^{+2468}_{-1711}$ & $7572^{+1805}_{-993}$ & $587^{+125}_{-79}$ & $1.0$ & $0.62$ \\[0.15cm] 

1375165725506487424 & ($231.849$, $36.034$) & ($57.9$, $55.86$) & $0.156 \pm 0.013$ & $-15.14 \pm 0.011$ & $-14.147 \pm 0.015$ & $-84.773 \pm 0.435$ & $11.19$ & $7789^{+832}_{-710}$ & $9429^{+497}_{-385}$ & $555^{+81}_{-69}$ & $1.0$ & $0.59$ \\[0.15cm] 

6257153998879080320 & ($226.415$, $-18.787$) & ($341.93$, $33.73$) & $0.101 \pm 0.015$ & $-9.219 \pm 0.018$ & $-9.861 \pm 0.015$ & $336.968 \pm 1.76$ & $12.15$ & $11315^{+2420}_{-1620}$ & $6977^{+1867}_{-1023}$ & $560^{+138}_{-86}$ & $0.96$ & $0.50$ \\[0.15cm]

\textbf{Extragalactic} \\[0.15cm]

6241679678390298880 & ($236.994$, $-20.512$) & ($349.46$, $26.01$) & $0.103 \pm 0.014$ & $-9.69 \pm 0.017$ & $-11.065 \pm 0.012$ & $-304.986 \pm 0.815$ & $12.59$ & $11410^{+2293}_{-1552}$ & $5691^{+1862}_{-1037}$ & $642^{+141}_{-89}$ & $0.22$ & $0.73$ \\[0.15cm] 

1204061267883975040 & ($238.111$, $20.197$) & ($33.45$, $48.2$) & $0.129 \pm 0.014$ & $-9.756 \pm 0.013$ & $-14.42 \pm 0.015$ & $-262.04 \pm 0.835$ & $12.92$ & $9713^{+1547}_{-1172}$ & $8518^{+1023}_{-660}$ & $604^{+124}_{-92}$ & $0.44$ & $0.72$ \\[0.15cm]

4182243409716233856 & ($296.209$, $-15.205$) & ($24.94$, $-18.54$) & $0.102 \pm 0.013$ & $-9.511 \pm 0.016$ & $-11.653 \pm 0.012$ & $-135.681 \pm 1.457$ & $12.63$ & $11279^{+1854}_{-1429}$ & $5972^{+1446}_{-929}$ & $613^{+131}_{-101}$ & $0.41$ & $0.61$ \\[0.15cm] 

1309092223502856576 & ($257.42$, $29.977$) & ($52.3$, $33.99$) & $0.084 \pm 0.01$ & $-5.097 \pm 0.011$ & $-6.631 \pm 0.012$ & $-485.643 \pm 0.523$ & $12.73$ & $14885^{+2634}_{-1991}$ & $12843^{+2277}_{-1616}$ & $524^{+87}_{-60}$ & $0.24$ & $0.56$ \\[0.15cm]

4200563090205883392 & ($290.3$, $-9.399$) & ($27.84$, $-10.85$) & $0.093 \pm 0.014$ & $-10.489 \pm 0.016$ & $-8.89 \pm 0.014$ & $-194.05 \pm 1.165$ & $12.31$ & $11553^{+2208}_{-1586}$ & $6038^{+1785}_{-1068}$ & $585^{+141}_{-100}$ & $0.33$ & $0.52$ \\[0.15cm]

\hline
\end{tabular}

\begin{tablenotes}

\item \emph{Note.} Distances and total velocities are quoted in terms of the median of the distribution, with uncertainties derived from the $16$th and $84$th percentiles.

\end{tablenotes}

\end{threeparttable}
\end{table}
\end{landscape}

In our search for the fastest stars in the Galaxy, we focused on stars with the highest probability of being unbound from the Galaxy. The absence of HVSs is not surprising, since even if their ejection rate is expected to be higher than the one of hyper-runaway stars \citep{brown15}, the bulk of the unbound HVS population is predicted to be too distant (and therefore too faint) to have a radial velocity measurement from {\Gaia} DR2 \citep{marchetti+18}. We cannot exclude the presence of bound HVSs in the sample considered in this work, as suggested by Fig. \ref{fig:rmin_E}.

The general validity of our results, in terms of observed and ejection velocity distributions and production mechanisms, is limited by the completeness function of {\Gaia} EDR3, and in particular be the selection function for the subset of sources with a radial velocity from {\Gaia} DR2. A number of studies have been focusing on determining the selection function of {\Gaia} DR2 \citep[e.g. \citet{Boubert+20I, Boubert+20II}, and][for the RVS sample]{Rybizki+20}, and these findings need to be included if one wants to infer rates or study the relative contribution of the different ejection mechanisms to the observed population of high velocity stars.

One possibility that we have not considered in this work is that our best candidates are actually \emph{bound} to the Milky Way. In this case, they could constitute the highest velocity tail of the velocity distribution of halo stars \citep[see][]{abadi+09}, as suggested by the observed colour-magnitude diagram shown in Fig. \ref{fig:HR} \citep[see also][]{hattori+18b}, and by spectroscopic follow-ups of a sample of fast stars in {\Gaia} DR2 presented in \citet{hawkins+18}. This might result either from the escape speed at their position being higher than the value adopted in this work \citep[see the discussion in][]{hattori+18b}, or because of inaccurate astrometry or radial velocities reported in {\Gaia} EDR3. In this work we have chosen to focus on the stars with the highest probabilities of being unbound to tackle the first issue. Results computed including the tentative parallax zero point correction introduced in \citet{Lindegren+20b} show that the number of unbound candidates decreases drastically, suggesting that these stars, if the parallax offset is confirmed, could actually be bound to the Milky Way (we refer the reader to Section \ref{sec:zp}). Furthermore, in Appendix \ref{appendix:stat_outliers} we estimate the fraction of possible statistical outliers contaminating our sample of high velocity star candidates, finding this number to be $\sim 40\%$. Future {\Gaia} data releases will improve even more the precision and the accuracy of the measurements thanks to the longer observational timeline. More precise data will also be essential to provide tighter constraints on the Galactic potential and to map the escape speed on a large range of distances from the Galactic Centre, allowing us to gain more confidence in the goodness of our candidates. 

Our knowledge on the fastest nearby stars will be revolutionized by the third data release of {\Gaia}, which is currently planned for the first half of 2022. {\Gaia} DR3 will provide new radial velocities for sources with $G_\mathrm{RVS} < 14$. A hundred of HVSs will be bright enough to have a radial velocity measurement \citep{marchetti+18}, even if their expected velocity distribution peaks at bound velocities, making their observational identification not trivial \citep{marchetti+18}. The bulk of the HVS population will not have a radial velocity from {\Gaia}, therefore a synergy with ground-based spectroscopic facilities will be essential to discover it. A larger sample of HVSs will be needed, for example, to provide tight constraints on the shape of the Milky Way dark matter halo \citep{Contigiani+19}. The synergy with upcoming spectroscopic surveys such as WEAVE \citep{weave} and 4MOST \citep{4MOST} will be essential to complement {\Gaia} precise astrometry, providing radial velocities and chemical abundances for tens of millions of stars. Furthermore, high resolution spectroscopy of the most interesting candidates would allow also to gain deeper insights into the ejection location of the fastest stars. For example, hyper-runaway stars ejected in the binary scenario are expected to exhibit evidence of polluting elements from the supernova ejecta \citep{przybilla+08, Pan+12}, and chemical tagging can be used to separate stars belonging to the disk, halo, and LMC populations \citep[e.g.][]{Hogg+16, hawkins+18}.

\section*{Acknowledgements}

The author thanks the anonymous referee for his/her comments, which greatly improved the quality of this manuscript.
The author thanks A. G. A. Brown and E. Zari for suggestions and advice on the use of {\Gaia} EDR3 astrometry, and F. Evans for useful discussions on the possible origin of extragalactic stars.
The author acknowledges an ESO fellowship.
This work has made use of data from the European Space Agency (ESA) mission
{\it Gaia} (\url{https://www.cosmos.esa.int/gaia}), processed by the {\it Gaia}
Data Processing and Analysis Consortium (DPAC,
\url{https://www.cosmos.esa.int/web/gaia/dpac/consortium}). Funding for the DPAC
has been provided by national institutions, in particular the institutions
participating in the {\it Gaia} Multilateral Agreement. This research made use of \textsc{Astropy}, a community-developed core \textsc{Python} package for Astronomy \citep{astropy}. All figures in the paper were produced using \textsc{matplotlib} \citep{matplotlib} and \textsc{Topcat} \citep{topcat}. This work would not have been possible without the countless hours put in by members of the open-source community all around the world.

\section*{Data Availability}

This work has made use of data from the European Space Agency (ESA) mission {\Gaia}, publicly available at the {\Gaia} archive. The catalogues derived in this work can be downloaded \href{https://sites.google.com/view/tmarchetti/research}{here}.

\bibliographystyle{mnras}
\bibliography{hvs.bib}

\appendix

\section{Comparison with other distance estimates}
\label{appendix:distances}

\begin{figure}
	\centering
 	\includegraphics[width=\columnwidth]{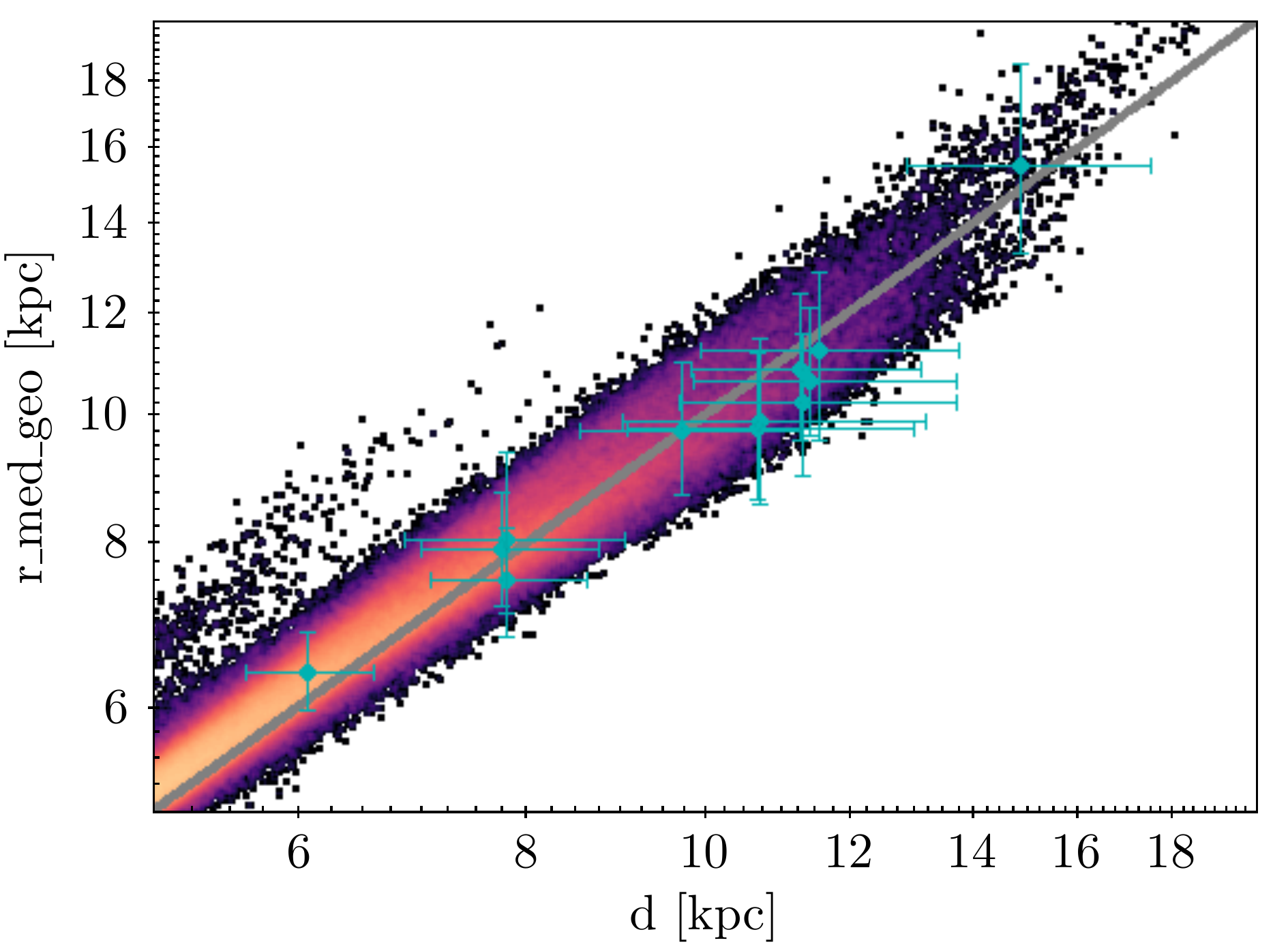}
 	\includegraphics[width=\columnwidth]{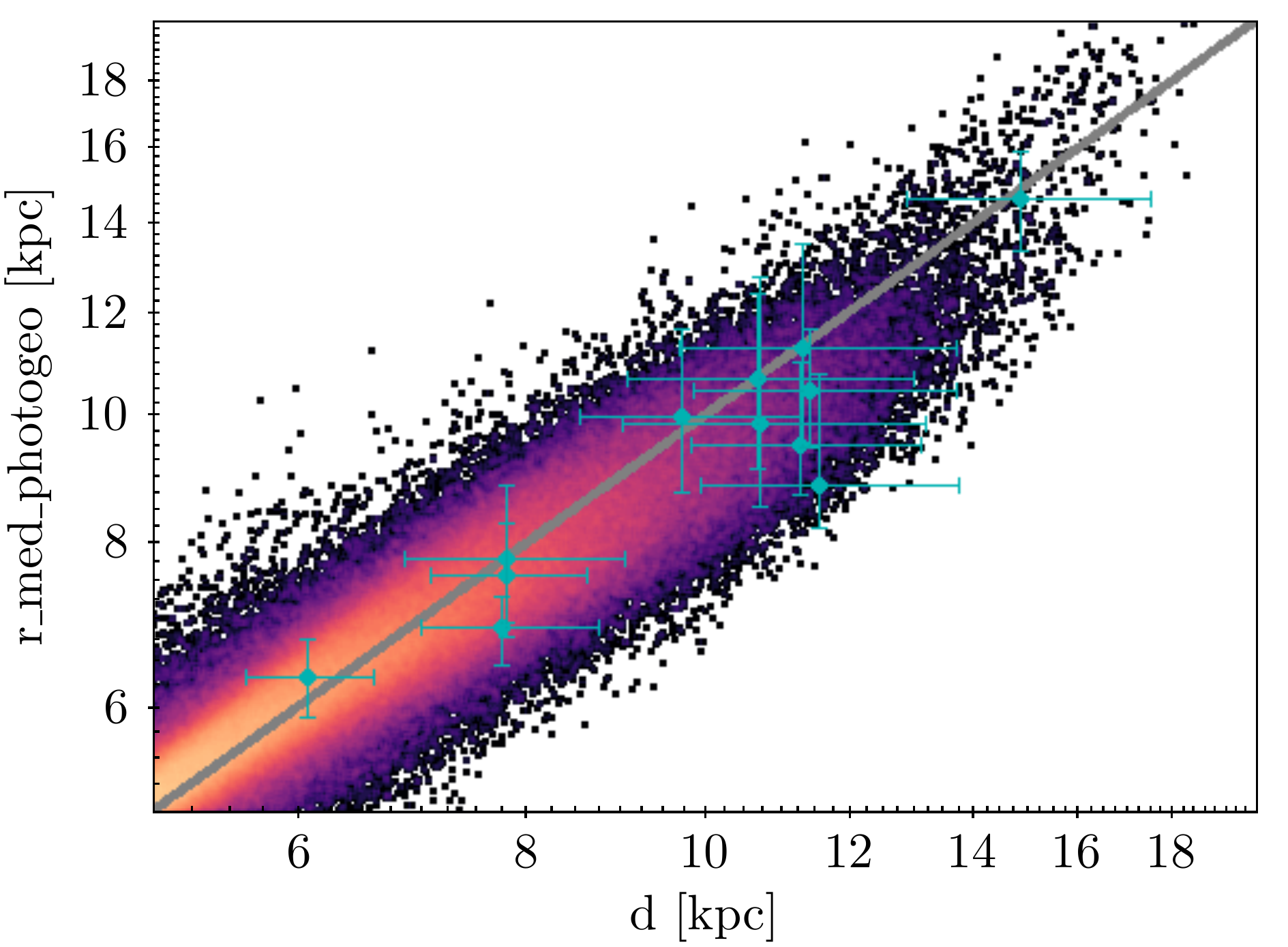}
	\caption{Distance comparisons for the subset of stars with $d >5$ kpc. The top panel (bottom panel) shows geometric (photogeometric) distances from \citet{bailer-jones+20} on the $y$ axis, while distances on the $x$ axis are derived in this work, including the effect of the parallax zero point. Blue points correspond to the 12 stars selected in Section \ref{sec:zp} with $P_\mathrm{ub}>0.5$.}
	\label{fig:dBJ}
\end{figure}

In this Appendix, we compare our distance estimates to the recent results of \citet{bailer-jones+20}. The authors derive distances for $\sim 1.47$ billion stars in {\Gaia} EDR3, using a Bayesian approach with priors constructed using realistic simulations of our Galaxy \citep{Rybizki+20b}. Two types of distances are determined: \emph{geometric} distances, inferred from the sky position, parallax, and parallax uncertainty; and \emph{photogeometric} distances, which depend also on the source magnitude and colour. Both distances are derived including the estimate of the parallax zero point introduced in \citet{Lindegren+20b}, therefore we extend our comparison only to the distances computed in Section \ref{sec:zp}. In Fig. \ref{fig:dBJ} we compare the distances computed in Section \ref{sec:zp} following the method outlined in Section \ref{sec:method}, to the distances inferred in \citet{bailer-jones+20}. In both panels, we can see how the majority of stars lie on the $y=x$ line, with a smaller spread in the case of purely geometric distances (top panel). Blue points refer to the $12$ stars selected in Section \ref{sec:zp} with $P_\mathrm{ub}>0.5$. Far all of these stars but one, {\Gaia} EDR3 4200563090205883392, our distance estimates are consistent, within the uncertainties, with those inferred in \citet{bailer-jones+20}. We note that the majority of stars have median distances $d$ higher then the values of geometric and photogeometric distances inferred in \citet{bailer-jones+20}. This can be explained by the fact that the values of the median of the distance prior used in the Bayesian approach in \citet{bailer-jones+20} range from $\sim 1$ kpc to $\sim 4.4$ kpc, lower than the distances determined by inverting the quoted {\Gaia} EDR3 parallaxes.

\section{Catalogue content}
\label{appendix:catalogue}

\begin{table*}
	
	\caption{Description of the columns of the catalogue with distances and velocities. Entries labelled $^1$ are taken directly from the {\Gaia} EDR3 catalogue, while entries labelled $^2$ are derived in this paper. We refer the interested reader to Section \ref{sec:method} for details on how these quantities are computed. For the entries labelled $^2$, the quoted values correspond to the median of the distribution, with lower and upper uncertainties computed, respectively, from the $16$th and the $84$th percentile. Column 60 is present only in the version of the catalogue which includes the effect of the parallax zero point.}
	
	\label{tab:catalogue}
	
	\begin{threeparttable}
	\setlength{\tabcolsep}{12pt}
		
		\begin{tabular}{llll}
		
			\hline		
			Column & Units & Name & Description \\
			\hline
			
			
			1 & - & source\_id & {\Gaia} EDR3 identifier$^1$ \\
			2 & deg & ra & Right ascension$^1$ \\
			3 & deg & dec & Declination$^1$ \\
			4 & mas & parallax & Parallax$^1$ \\
			5 & mas & e\_parallax & Standard uncertainty in parallax$^1$ \\
			6 & mas yr$^{-1}$ & pmra & Proper motion in right ascension$^1$ \\
			7 & mas yr$^{-1}$ & e\_pmra & Standard uncertainty in proper motion in right ascension$^1$ \\
			8 & mas yr$^{-1}$ & pmdec & Proper motion in declination$^1$ \\
			9 & mas yr$^{-1}$ & e\_pmdec & Standard uncertainty in proper motion in declination$^1$ \\
			10 & - & corr\_parallax\_pmra & Correlation between parallax and proper motion in right ascension$^1$\\
			11 & - & corr\_parallax\_pmdec & Correlation between parallax and proper motion in declination$^1$\\
			12 & - & corr\_pmra\_pmdec & Correlation between proper motion in right ascension and proper motion in declination$^1$\\
			13 & km s$^{-1}$ & vrad & {\Gaia} DR2 Radial velocity$^1$ \\
			14 & km s$^{-1}$ & e\_vrad & {\Gaia} DR2 Radial velocity error$^1$ \\
			15 & mag & GMag & G-band mean magnitude$^1$ \\
			16 & mag & BPMag & BP-band mean magnitude$^1$ \\
			17 & mag & RPMag & RP-band mean magnitude$^1$ \\
			18 & - & \textsc{ruwe} & Renormalised Unit Weight Error$^1$ \\
			19 & - & \textsc{rv\_nb\_transits} & Number of epochs used in {\Gaia} DR2 to determine the radial velocity$^1$ \\
			
			
			20 & pc & dist & Distance estimate$^2$ \\
			21 & pc & el\_dist & Lower uncertainty on distance$^2$ \\
			22 & pc & eu\_dist & Upper uncertainty on distance$^2$ \\
			23 & pc & rGC & Spherical Galactocentric radius$^2$ \\
			24 & pc & el\_rGC & Lower uncertainty on spherical Galactocentric radius$^2$ \\
			25 & pc & eu\_rGC & Upper uncertainty on spherical Galactocentric radius$^2$ \\
			26 & pc & RGC & Cylindrical Galactocentric radius$^2$ \\
			27 & pc & el\_RGC & Lower uncertainty on cylindrical Galactocentric radius$^2$ \\
			28 & pc & eu\_RGC & Upper uncertainty on cylindrical Galactocentric radius$^2$ \\
			29 & pc & xGC & Cartesian Galactocentric $x$-coordinate$^2$ \\
			30 & pc & el\_xGC & Lower uncertainty on Cartesian Galactocentric $x$-coordinate$^2$ \\
			31 & pc & eu\_xGC & Upper uncertainty on Cartesian Galactocentric $x$-coordinate$^2$ \\		
			32 & pc & yGC & Cartesian Galactocentric $y$-coordinate$^2$ \\
			33 & pc & el\_yGC & Lower uncertainty on Cartesian Galactocentric $y$-coordinate$^2$ \\
			34 & pc & eu\_yGC & Upper uncertainty on Cartesian Galactocentric $y$-coordinate$^2$ \\
			35 & pc & zGC & Cartesian Galactocentric $z$-coordinate$^2$ \\
			36 & pc & el\_zGC & Lower uncertainty on Cartesian Galactocentric $z$-coordinate$^2$ \\
			37 & pc & eu\_zGC & Upper uncertainty on Cartesian Galactocentric $z$-coordinate$^2$ \\
		
			
			38 & km s$^{-1}$ & U & Cartesian Galactocentric $x$-velocity$^2$ \\
			39 & km s$^{-1}$ & el\_U & Lower uncertainty on Cartesian Galactocentric $x$-velocity$^2$ \\
			40 & km s$^{-1}$ & eu\_U & Upper uncertainty on Cartesian Galactocentric $x$-velocity$^2$ \\
			41 & km s$^{-1}$ & V & Cartesian Galactocentric $y$-velocity$^2$ \\
			42 & km s$^{-1}$ & el\_V & Lower uncertainty on Cartesian Galactocentric $y$-velocity$^2$ \\
			43 & km s$^{-1}$ & eu\_V & Upper uncertainty on Cartesian Galactocentric $y$-velocity$^2$ \\
			44 & km s$^{-1}$ & W & Cartesian Galactocentric $z$-velocity$^2$ \\
			45 & km s$^{-1}$ & el\_W & Lower uncertainty on Cartesian Galactocentric $z$-velocity$^2$ \\
			46 & km s$^{-1}$ & eu\_W & Upper uncertainty on Cartesian Galactocentric $z$-velocity$^2$ \\
			47 & km s$^{-1}$ & UW & Cartesian Galactocentric $xz$-velocity$^2$ \\
			48 & km s$^{-1}$ & el\_UW & Lower uncertainty on Cartesian Galactocentric $xz$-velocity$^2$ \\
			49 & km s$^{-1}$ & eu\_UW & Upper uncertainty on Cartesian Galactocentric $xz$-velocity$^2$ \\ 	
			50 & km s$^{-1}$ & vR & Cylindrical Galactocentric $R$-velocity$^2$ \\
			51 & km s$^{-1}$ & el\_vR & Lower uncertainty on cylindrical Galactocentric $R$-velocity$^2$ \\
			52 & km s$^{-1}$ & eu\_vR & Upper uncertainty on cylindrical Galactocentric $R$-velocity$^2$ \\
			53 & km s$^{-1}$ & vphi & Cylindrical Galactocentric azimuthal velocity$^2$ \\
			54 & km s$^{-1}$ & el\_vphi & Lower uncertainty on cylindrical Galactocentric azimuthal velocity$^2$ \\
			55 & km s$^{-1}$ & eu\_vphi & Upper uncertainty on cylindrical Galactocentric azimuthal velocity$^2$ \\			
			
			56 & km s$^{-1}$ & vtot & Total velocity in the Galactic rest-frame$^2$ \\
			57 & km s$^{-1}$ & el\_vtot & Lower uncertainty on total velocity in the Galactic rest-frame$^2$ \\
			58 & km s$^{-1}$ & eu\_vtot & Upper uncertainty on total velocity in the Galactic rest-frame$^2$ \\
			
			59 & - & P\_ub & Probability of being unbound from the Galaxy$^2$ \\
			
			60 & mas & parallax\_zp & Parallax zero point, estimated as in \citet{Lindegren+20b} \\

			\hline
		
		\end{tabular}
		
	\end{threeparttable}
	
\end{table*}

A description of the columns provided in the catalogue containing derived distances and velocities for the 6969738 stars with {\Gaia} EDR3 precise astrometry ($\varpi > 0$ and $\sigma_\varpi/\varpi < 0.2$) and {\Gaia} DR2 radial velocities is presented in Table \ref{tab:catalogue}. The catalogue can be downloaded as a single FITS file \href{https://sites.google.com/view/tmarchetti/research}{here}. 

At the same web address, it is also possible to download the catalogue constructed subtracting the estimated parallax zero point to the parallax of each source, as discussed in Section \ref{sec:zp}. In this case, the quoted parallax in the fourth column of the catalogue is once again the one provided by {\Gaia} EDR3, and the estimated parallax zero point $\varpi_\mathrm{zp}$ for each source is given as an additional last column.

\section{Revisiting the fastest stars in {\Gaia} DR2} 
\label{appendix:highV_DR2}

We have already remarked the small overlap between the new unbound star candidates found in {\Gaia} EDR3 and the best candidates found in {\Gaia} DR2 by \citetalias{Marchetti+19}. In this Appendix, we revisit the $20$ stars with $P_\mathrm{ub}>0.8$ presented in \citetalias{Marchetti+19}, and we discuss the reasons why they are no longer classified as high velocity stars when using the new and more precise {\Gaia} EDR3 astrometry.

We verify that the {\Gaia} DR2 identifier has not changed in {\Gaia} EDR3 for all the objects. {\Gaia} EDR3 1383279090527227264, the only candidate from {\Gaia} DR2 which is selected by our updated search in this paper (see discussion in Section \ref{sec:galactic} and Table \ref{tab:candidates}), will not be considered here.

\begin{itemize}

    \item 1 star, {\Gaia} EDR3 5932173855446728064, does not have a {\Gaia} DR2 radial velocity in {\Gaia} EDR3. This star was the most solid unbound candidate found by \citetalias{Marchetti+19} and \cite{Bromley+18}, with an extremely precise total velocity of $\sim 750$ {\kms}. This high velocity was due to the radial velocity measured by {\Gaia} DR2, equal to $-614.5$ {\kms}. Subsequent spectroscopic follow-ups presented in \citet{Boubert+19} reported a radial velocity of $-56.5$ {\kms}, remarkably different from the one measured by {\Gaia}. \citet{Boubert+19} explain this large discrepancy between the measurements as due to the contamination from a nearby bright star to the spectrum of the source collected by the RVS. This wrong radial velocity from {\Gaia} DR2 was therefore excluded from {\Gaia} EDR3.
    
    \item 7 stars do not satisfy the parallax cuts introduced in Section \ref{sec:method}. These objects are {\Gaia} EDR3 1396963577886583296, {\Gaia} EDR3 5593107043671135744, {\Gaia} EDR3 5546986344820400512, {\Gaia} EDR3 5257182876777912448, {\Gaia} EDR3 5831614858352694400, {\Gaia} EDR3 5830109386395388544 and {\Gaia} EDR3 4073247619504712192. All but one of these stars were classified as \emph{extragalactic stars} in \citetalias{Marchetti+19}. These stars have values of $f=\sigma_\varpi/\varpi$ in the range $[0.21, 1.08]$ in {\Gaia} EDR3, making their distance determination highly uncertain. Spectro-photometric distances are needed in order to confirm or reject their nature as velocity outliers.
    
    \item 1 star, {\Gaia} EDR3 1990547230937629696, has \textsc{ruwe} $= 1.49$, and is therefore excluded according to the cuts introduced in Section \ref{sec:cuts}. Interestingly, the reported \textsc{ruwe} in {\Gaia} DR2 is $1.17$. The increase in \textsc{ruwe} from {\Gaia} DR2 to {\Gaia} EDR3 could be a hint that this star is an unresolved binary system \citep[e.g.][]{Belokurov+20}.
    
    \item The other 10 stars identified in \citetalias{Marchetti+19} have $P_\mathrm{ub} < 0.5$ when using the new astrometry from {\Gaia} EDR3. The fastest object among these is {\Gaia} EDR3 5935868592404029184, with $v_\mathrm{GC} = 613$ {\kms}, and $P_\mathrm{ub} = 0.44$.
    
\end{itemize}

\section{Estimating the fraction of statistical outliers}
\label{appendix:stat_outliers}

Since we are looking for the fastest objects in a catalogue of $\sim 7$ million stars, in this Appendix we quantify the fraction of high velocity star candidates identified in this work that are likely to be statistical outliers. We sample the astrometry of each star by drawing a random realization of its astrometry vector $\mathbf{m^\prime}$ from a multi-variate Gaussian distribution centered on the observed mean vector $\mathbf{m}$ (see Equation \ref{eq:mean}) and with covariance matrix given by Equation \eqref{eq:covmatr}. We then repeat the same analysis described in Section \ref{sec:method}, drawing $5000$ MC realizations of the astrometry of each source from a multi-variate Gaussian distribution centered on $\mathbf{m^\prime}$ with covariance matrix given by Equation \eqref{eq:covmatr}, and then we derive distances and total velocities. We repeat this approach $12$ times, and we check how many of the best candidates identified in Section \ref{sec:high_vel} are retrieved when applying the quality cuts introduced in Section \ref{sec:cuts} and $P_\mathrm{ub}>0.8$. We find that, on average, we recover $10$ of the $17$ stars listed in Table \ref{tab:candidates}, suggesting that up to $\sim 40\%$ of the fastest stars in the sample could be statistical outliers. The more precise astrometry from future data releases of {\Gaia} will be crucial to identify the most robust candidates.

\bsp

\label{lastpage}

\end{document}